\documentclass[%
 reprint,%
 amssymb, amsmath,%
 aip,rsi,%
]{revtex4-1}

\usepackage{graphicx}
\usepackage{docs}%
\usepackage{bm}%
\usepackage[colorlinks=true,linkcolor=blue]{hyperref}%
\expandafter\ifx\csname package@font\endcsname\relax\else
 \expandafter\expandafter
 \expandafter\usepackage
 \expandafter\expandafter
 \expandafter{\csname package@font\endcsname}%
\fi
\newcommand*{\I}[1]{\mathrm{i}}

\newcommand*{\pumpaxis}[0]{y}
\newcommand*{\probeaxis}[0]{x}
\newcommand*{\sensaxis}[0]{z}

\begin{document}

\title{Instrumentation for nuclear magnetic resonance in zero and ultralow magnetic field}%

\author{Michael C. D. Tayler}
\email{mcdt2@cam.ac.uk}
\affiliation{Department of Physics, UC Berkeley, California, United States}
\affiliation{Department of Chemical Engineering and Biotechnology, Cambridge University, United Kingdom}
\author{Thomas Theis}
\affiliation{Department of Chemistry, Duke University, Durham, North Carolina, United States}
\author{Tobias F. Sjolander}
\affiliation{College of Chemistry, UC Berkeley, California, United States}
\author{John W. Blanchard}
\affiliation{Helmholtz Institut Mainz, 55099 Mainz, Germany}
\author{Arne Kentner}
\affiliation{RWTH Aachen, Germany}
\author{Szymon Pustelny}
\affiliation{Institute of Physics, Jagiellonian University, 30-348 Krak\'{o}w, Poland}
\author{Alexander Pines}
\affiliation{College of Chemistry, UC Berkeley, California, United States}
\author{Dmitry Budker}
\affiliation{Department of Physics, UC Berkeley, California, United States}
\affiliation{Nuclear Sciences Division, Lawrence Berkeley National Laboratory, California, United States}
\affiliation{Helmholtz Institut Mainz, Johannes Gutenberg Universit\"{a}t, Mainz, Germany}

\date{May 9$^{\mathrm{th}}$, 2017}%

\maketitle

\section*{Abstract}
We review instrumentation for nuclear magnetic resonance (NMR) in zero and ultra-low magnetic field (ZULF, below 0.1 $\mu$T) where detection is based on a low-cost, non-cryogenic, spin-exchange relaxation free (SERF) $^{87}$Rb atomic magnetometer.  The typical sensitivity is 20-30 fT/Hz$^{1/2}$ for signal frequencies below 1 kHz and NMR linewidths range from Hz all the way down to tens of mHz.  These features enable precision measurements of chemically informative nuclear spin-spin couplings as well as nuclear spin precession in ultra-low magnetic fields.    

\tableofcontents

\section{Introduction}
Nuclear magnetic resonance (NMR) signals are commonly detected using inductive coupling,\cite{Hill2011emagres} where the sample is placed in a magnetic field and the Larmor precession of the atomic nuclei induces an electromotive force (emf) in a nearby circuit.  Inductive detection favors high magnetic fields because the emf is proportional to the magnetic field.  Additionally, a higher field offers a larger equilibrium nuclear spin polarization\cite{Iriguchi1993JAP73,Sykora2005ebyte} to improve sensitivity.  Overall, signal to noise ratio (SNR) scales with $B^{7/4}$, because signal is proportional to both polarization and emf, while noise scales as $B^{1/4}$.\cite{Abragambook1961,Iriguchi1993JAP73,Hoult1976JMR24,Hoult1979JMR34,Hoult2007EMR}  High fields also increase chemical shift dispersion for experiments addressing specific nuclear spins and/or chemical environments.\cite{Freeman1991CR91}  At present a top-of-the-range high-field NMR instrument with a 10-20 T superconducting magnet (approaching the GHz frequency regime) and a low-noise or cryo-cooled induction circuit\cite{Styles1984JMR60} can be used for analyses of chemical samples down to the picomole level and of sample volumes around $1-10$ $\mu$L.\cite{Martin2005ARNMR56,Lacey1999CR99}  

However, there are many applications of NMR for which there is a preference for detecting signals at low field: (1) The study of samples containing phase boundaries and therefore large variations in magnetic susceptibility, e.g.\ mixed phases and porous materials, can be challenging in high field.  Spectral peaks are broad because of the induced inhomogeneity in the NMR magnetic field.  In contrast, towards zero field, the broadening becomes small compared to the natural linewidth of the resonances.\cite{McDermott2002Science295}  (2) The NMR spectra of orientationally disordered samples, e.g.\ glassy or powdered solids, are broad at high field due to the truncation of spin interactions by the imposed symmetry of the magnetic field.  In low fields, where interactions with the external field are weak compared to spin-spin couplings, there is no truncation and spectra are sharp.\cite{Thayer1987ACR20,Zax1985JCP83,BlanchardPRL2015}  (3) For samples enclosed in metal containers, NMR signals can be strongly attenutated due to the skin-depth effect, which scales with the inverse square root of frequency.  At 10 MHz in copper or aluminum the penetration depth is around 20 microns; in contrast, at 1 kHz it is around 2 mm, deep enough to allow signals to pass through the walls of, say, a soda can.\cite{Mossle2006JMR179,Freedman2014RSI85} (4) At zero field, spin coherences in liquids can persist for many tens of seconds, corresponding to ultra-narrow linewidths on the order of mHz.  Such narrow lines can be used for chemical fingerprinting and precise measurement of spin-spin couplings.\cite{Emondts2014PRL112,Ledbetter2009JMR199,Butler2013JCP1382,Blanchard2013JACS,Theis2013CPL580,Blanchard2016emagres,Ledbetter2011PRL107,Ledbetter2013PT66}   Combined with the ease of obtaining highly homogeneous low fields, ZULF represents a facile route to chemically resolved NMR.

ZULF NMR demands instrumentation that is different to that from conventional NMR since inductive detection of the low-frequency signals is extremely time consuming.\cite{Lenz1990IEEEProc78}  In this paper, we review instrumentation for non-inductive NMR detection based on alkali-atom magnetometers.\cite{Kominis2003Nature422,Savukov2005PRL,Savukov2007JMR185,Ledbetter2008PRA77,Bevilacqua2009JMR201}  These magnetometers measure the magnitude of magnetic field emanating from the NMR sample via the ground-state precession of angular momenta in vaporized alkali metal atoms, e.g.\ K, Rb or Cs.\cite{Tiporlini2013TSW,Budker2007NatPhys3,Budker2002RMP74,BudkerOpticalMagnetometry,Seltzerthesis}  Alkali magnetometers are most sensitive at low fields owing to the alkali vapor becoming spin-exchange relaxation free (SERF) at fields below $\sim100$ nT.\cite{Happer1973PRL31,Happer1977PRA16,Savukov2005PRA71,Allred2002PRL89}  The fundamental sensitivity of SERF magnetometers is comparable to that of superconducting quantum interference devices (SQUIDs),\cite{Greenberg1998RMP70,Friedman1986RSI} which for the past 25 years had been the only viable means of directly detecting ZULF NMR signals \cite{McDermott2002Science295,Qiu2009IEEETAS22,Augustine1998SSNMR11} or ZULF magnetic resonance images\cite{Fan1989IEEETM2,KrausJrbook,Espy2005IEEETAS15} (as a guideline, SQUID sensitivity surpasses that of an inductive coil for frequencies below 1-10 MHz, taking spin polarization, sample volume and all other factors as equal\cite{Matlashov2011IEEEApplSupercon21}).  Like for a SQUID, the sensitivity of an atomic magnetometer is broadly independent of detection frequency.  However, the bandwidth is not as large.  Depending on design, SQUIDs can detect across several hundreds of kHz, but atomic magnetometers typically reach only a few hundreds of Hz down to a few Hz in the most sensitive SERF regime.  On the other hand, ultra-low-field NMR spectra of liquid-state samples containing only $^{1}$H, $^{13}$C, $^{19}$F or $^{15}$N nuclei usually span less than 1 kHz, so in this application the limitation is not a major one.  A highly attractive feature of atomic magnetometers with respect to SQUIDs is that they do not require cryogenic cooling -- in fact, they operate at or above room temperature -- and are therefore significantly less complex and costly.  Other features, for example a much smaller magnetic shield needed to obtain an ultralow field environment around the sensor, enable the atomic magnetometer apparatus to be compact and to fit on a laboratory benchtop while the shielding and other hardware associated with SQUIDs is presently much larger.

Over the past decade SERF atomic magnetometers were applied, down to chip-scale,\cite{Ledbetter2008PNAS105,Shah2007NatPhot1,Martinez2014NatCommun5} in many situations where earth's field NMR has normally been used, e.g.: hyperpolarized gas imaging,\cite{Savukov2009JMR199} spin relaxation and molecular diffusion measurement in the absence of resolved chemical shifts,\cite{Ganssle2014ANIE53} and the quantification of net spin order in strongly polarized samples.\cite{Yashchuk2004PRL93}  They have also been extensively used to obtain NMR spectra with ultrahigh (mHz) resolution of the scalar spin-spin and dipolar couplings, where the measurement precision is typically an order of magnitude higher than achieved using high-field NMR.\cite{Blanchard2016emagres,Wilzewski2017} Technical development of the magnetometers is presently in a stage of rapid advance due also to their use in many other scientific areas including biomagnetism\cite{Schwindt2010Sandia} and fundamental physics.\cite{Baker2006PRL97}

In the following section we describe the appearance of NMR signals in ultra-low and zero magnetic field.  We then explain how these are measured by describing our apparatus in the chronological order of the experiment, which involves (1) polarization, (2) encoding and (3) detection of the nuclear spins.  The atomic magnetometer used in this work had a sensitivity of $3\times 10^{-14}$ T Hz$^{-1/2}$, which is a compromise between bandwidth, sensitivity and size as explained below.  Details are given for building, calibrating and then operating the instrument to perform NMR measurements.  We conclude with a discussion of future development and possible improvements.

\section{Spin dynamics at ultra-low field} 
\label{sec:NMRspindynamics}
The NMR spectrum is obtained by Fourier transformation of a time-domain signal from the magnetometer whose amplitude is proportional to the total nuclear magnetization of the NMR sample.  We define the total magnetization along $z$ as our measurable.  This is a convenient yet arbitrary choice as is detailed further in Section \ref{sec:detectionintro}.  With this choice the signal is given as 
\begin{eqnarray}
\mathrm{signal} (t) &\propto& \langle M_z \rangle(t) \,\, \equiv \,\, \sum_{j}{h\gamma_j \langle I_{z}^{(j)} \rangle /2\pi}.
\end{eqnarray}
In the above expression, $\langle M_z \rangle$ is the ensemble-averaged magnetization of the sample along the $z$-axis, which equates to the total expectation value of angular momentum, $\langle I_{z}^{(j)} \rangle$, for each nucleus $j$ in the system multiplied by its gyromagnetic ratio, $\gamma_j$.  

For convenience we set $h = 1$ and measure energies in Hz.  The time dependence of $\langle M_z \rangle$ is computed in the density matrix formalism from the trace over the operator product $M_z\rho(t)$, where $\rho(t)$ is the nuclear spin density matrix operator for the sample.  The nuclear spin Hamiltonian, $H(t)$, determines the time-dependence of $\rho(t)$ by propagation.  In the case where $H$ is time-independent and spin relaxation is ignored, the result is
\begin{eqnarray}
\langle M_z \rangle(t) =  \mathrm{Tr}(M_z^\dagger \exp[-2\pi \mathrm{i} H t]\rho(0)\exp[+2\pi\mathrm{i} H t]).
\label{eq:Mzrho}
\end{eqnarray}

We shall give three basic instances of this equation to illustrate the NMR phenomenon at ultra-low magnetic field.  The first is Larmor precession of magnetically equivalent nuclei with spin quantum number $I=1/2$ in the presence of an applied field $\bm{B}=\{B_x, B_y, B_z\}$, (e.g.\ the $^1$H nuclei in a sample of water, oil or ethanol).  In this instance, the Hamiltonian is $H= (\gamma/2 \pi) \bm{I}^{(j)} \cdot \bm{B}$, where $\bm{I}^{(j)} = \{ I_{x}^{(j)}, I_{y}^{(j)}, I_{z}^{(j)} \}$ and the exponential parts of Equation \ref{eq:Mzrho} correspond to a rotation operator $R = \exp[-\mathrm{i} \gamma (\bm{I}^{(j)} \cdot \bm{B}) t]$; the spins precess about the axis of the applied field at an angular velocity $\gamma \bm{B}$ and $\rho(t) = R \rho(0)R^{-1}$.  Thus, if $^1$H magnetization is initially prepared along the $z$ axis, and a magnetic field is applied along the $x$ axis, ($|\bm{B}|=B_x$, $B_y=0$, $B_z=0$), the result will be $\langle M_z \rangle(t) = \langle M_z \rangle(0) \cos(\gamma_\mathrm{H} B_x t)$.  This NMR experiment is itself a form of magnetometry as the value of $B_x$ can be deduced from the measured signal frequency.  

\begin{figure}%
\includegraphics[width=\columnwidth]{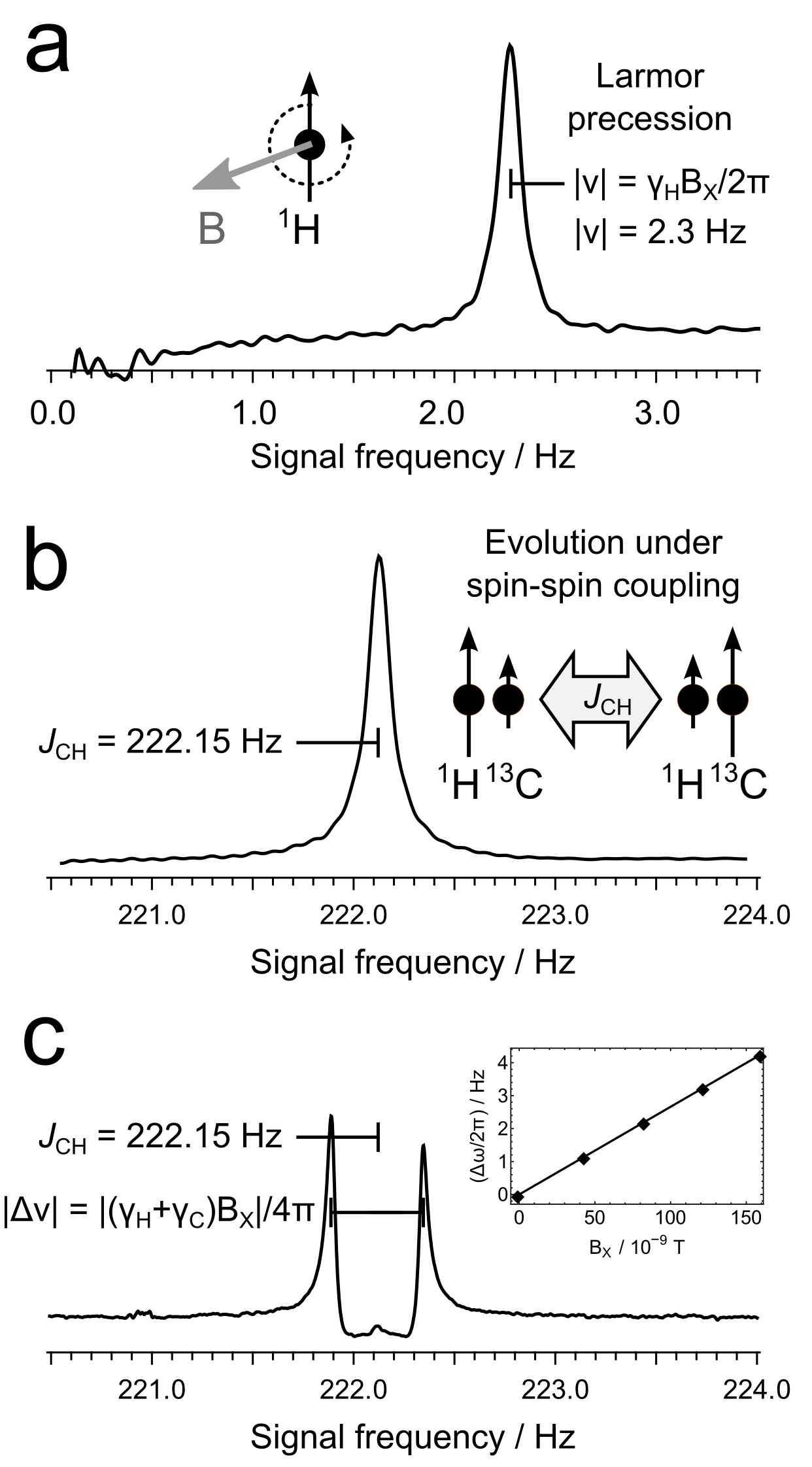}%
\caption{Examples of low-frequency NMR signals detected at ultra-low field: (a) Larmor precession of water in a bias field $\sim 54$ nT field applied perpendicular to the detection axis ($x$ and $z$ axes respectively); (b) measurement of the carbon-hydrogen spin-spin scalar coupling in [$^{13}$C]-formic acid in the absence of a bias field (``zero field''); (c) near-zero-field NMR signals of [$^{13}$C]-formic acid where the field direction is the same as that applied in (a).  The Zeeman interaction is perturbative relative to the spin-spin coupling, causing splittings that are symmetrical about the zero-field transition.}%
\label{fig:waterFA}
\end{figure}

An example of the above is shown in Figure \ref{fig:waterFA}a.  A spectrum was recorded using the atomic magnetometer for a $100$ $\mu$L sample of tap water in a field along the $\pumpaxis$ axis, produced by a current of 0.3 mA applied to a wire solenoid surrounding the sample; the value of the gyromagnetic ratio is $\gamma_\mathrm{H}$ = $2.675\times 10^8$ rad s$^{-1}$ T$^{-1}$, therefore the signal at $|\gamma_\mathrm{H} B_\pumpaxis /2\pi|$ = 2.3 Hz corresponds to precession in the field of the coil, around $B_\pumpaxis = 5.40\times 10^{-8}$ T.  The initial condition in the experiment was polarization of the spins along the $\sensaxis$ axis at a level of approximately 1 ppm, achieved by placing the sample in the field of a permanent magnet (2 T) located approximately 20 cm away from the magnetometer followed by rapid shuttling into the low-field region.  Further details on this protocol are given in section \ref{sec:zeemanpolarization}.

In this case, the magnetometer is sensitive to the projection of field along a single axis and the sign of the precession frequency cannot be determined.  To determine the sign, quadrature detection would be needed, which could be realized using two magnetometers, the second sensitive to fields along the $x$ or $y$ axis.

Nuclear spin-coherence lifetime and magnetic field homogeneity determine the precision to which $B_x$ can be measured, since both of these parameters influence the linewidth of the spectral peak.  For the water sample in Figure \ref{fig:waterFA}a the lifetime ($\approx4$ s) is the dominant factor.  More precise measurements of field magnitude $|\bm{B}|$ can be achieved using other nuclei, e.g.\ $^{3}$He or $^{129}$Xe.  Although these nuclei have relatively low gyromagnetic ratios ($\gamma_\mathrm{3He}\approx0.76 \gamma_\mathrm{H}$ and $\gamma_\mathrm{129Xe} \approx0.28 \gamma_\mathrm{H}$ respectively), they can sustain coherence lifetimes on the order of hours due to the chemical inertness of the atoms.  Strong signals can be obtained through spin-exchange optical-pumping techniques.\cite{Maul2016RSI87,Kornack2005PRL95}  The measurements of precession can be precise enough for navigation-grade gyroscopes.\cite{Donley2010IEEEsensCon}

The second instance of Equation \ref{eq:Mzrho} concerns systems of coupled nuclear spins.  The NMR spectrum of a liquid is also a function of the indirect dipole-dipole couplings between the nuclear spins, which are mediated by the molecular electrons and described by a term in the intramolecular Hamiltonian $H_J = \sum_{j,k>j}{J_{jk}\bm{I}^{(j)}\cdot \bm{I}^{(k)}}$.  A simple system illustrating this is [$^{13}$C]-formic acid (H$^{13}$COOH) in zero field.  The spin system is assumed to comprise the spin-1/2 pair involving the $^{13}$C and $^1$H nuclei on the formyl group (the acidic $^1$H undergoes rapid chemical exchange and can be ignored). This leads to $H=H_J = J_\mathrm{CH} (I_{x}^\mathrm{H}I_{x}^\mathrm{C} + I_{y}^\mathrm{H}I_{y}^\mathrm{C} + I_{z}^\mathrm{H}I_{z}^\mathrm{C})$.  We have defined ``zero field'' as where the Larmor frequencies are much smaller than both the inverse of the spin coherence time and the spin-spin couplings, so the Zeeman interaction is dropped from the Hamiltonian.

The coupling constant $J_{\mathrm{CH}} = 222.15$ Hz results in a peak in the zero-field spectrum, as shown in Figure \ref{fig:waterFA}b.  In a similar way to our example of $^1$H Larmor precession in water, the signal arises when the starting nuclear spin density operator $\rho(0)$ does not commute with $H$.  One such operator is $\rho(0) = k (I_{z}^\mathrm{H} - I_{z}^\mathrm{C})$, where $k$ is a constant. The set of operators $A_1 = (I_{x}^\mathrm{H}I_{x}^\mathrm{C} + I_{y}^\mathrm{H}I_{y}^\mathrm{C})$, $A_2 = (I_{z}^\mathrm{H} - I_{z}^\mathrm{C})/2$ and $A_3 = (I_{x}^\mathrm{H}I_{y}^\mathrm{C} - I_{y}^\mathrm{H}I_{x}^\mathrm{C})$ obey the commutation relationship $[A_1,A_2] = \mathrm{i} A_3$ plus cyclic permutations $[A_2,A_3] = \mathrm{i} A_1$ and $[A_3,A_1] = \mathrm{i} A_2$ and in addition $[A_1, A_4] = [A_2, A_4] = [A_3, A_4] = 0$ with $A_4=I_{z}^\mathrm{H}I_{z}^\mathrm{C}$.  These allow one to use the standard result $\exp[\mathrm{i} \theta (A_1+A_4)] A_2 \exp[-\mathrm{i} \theta (A_1+A_4)] = A_2\cos\theta + A_3\sin\theta$ to evaluate Equation \ref{eq:Mzrho} given the initial condition and reach the result $\rho(t) = k [(I_{z}^\mathrm{H} - I_{z}^\mathrm{C}) \cos(2 \pi J_\mathrm{CH} t) + (2I_{x}^\mathrm{H}I_{y}^\mathrm{C} - 2I_{y}^\mathrm{H}I_{x}^\mathrm{C})\sin(2 \pi J_\mathrm{CH} t)]$.  It is then deduced that $\langle M_z \rangle(t) = \langle M_z \rangle(0) \cos(2\pi J_\mathrm{CH} t)$; the magnetization oscillates at the frequency $J_\mathrm{CH}$.  Details for the preparation of $\rho(0)$ may be found in sections \ref{sec:zeemanpolarization} and \ref{sec:encoding}. 
 
More generally the appearance of the ZULFs NMR spectra can be determined by expanding Equation \ref{eq:Mzrho} in terms of matrix elements between the normalized eigenstates $\{|\psi_j\rangle\}$ of $H$:
\begin{eqnarray}
\langle M_z \rangle(t) &=& \sum_{k>j,j} \langle{\psi_j | \rho(0) | \psi_k}\rangle \nonumber \\
&& \qquad \qquad \times \langle{\psi_j | M_z | \psi_k}\rangle \exp(-2\pi \mathrm{i} \nu_{jk} t).
\label{eq:selectionrule}
\end{eqnarray}
The exponents contain the eigenfrequencies, i.e.\ energy differences between eigenstates, denoted by $\nu_{jk} \equiv (\langle{\psi_j | H | \psi_j}\rangle - \langle{\psi_k | H | \psi_k}\rangle)$.  The appearance of the NMR spectra for more than two coupled spins is extensively discussed in the literature.\cite{Butler2013JCP1382,Theis2013CPL580,Blanchard2016emagres}  As a general rule, oscillatory signals are not detectable if the detection operator $M_z$ commutes with $H$.  Any two matrix operators that commute have the same eigenstates.  Thus, the matrix elements $\langle{\psi_j | M_z | \psi_k}\rangle$ are zero for $| \psi_j\rangle \neq | \psi_k\rangle$ because $|\psi_k\rangle$ and $|\psi_k\rangle$ are eigenstates of $M_z$ and are orthonormal.  Although matrix elements $\langle{\psi_j | M_z | \psi_j}\rangle$ can be nonzero, the transition frequency is zero and so the rule holds.  For a system at zero field, where all of the spins have the same gyromagnetic ratio, $M_z$ and $H_J$ always commute. It can be concluded that detection of the zero-field spectrum therefore requires coupling between more than one spin species in the system.  

While the frequencies of the ultralow-field NMR signals can be predicted exactly using the eigenvalue-eigenvector approach (Equation \ref{eq:selectionrule}), for fields below $0.1$ $\mu$T the Zeeman interaction is small enough to be treated as a first-order perturbation to the zero-field eigensystem.\cite{Ledbetter2011PRL107}  When a bias field is applied along the $x$ or $y$ axes, transitions in the NMR spectrum appear split into $n=2(2F+1)$ equal-spaced components corresponding to $\Delta m_F = \pm 1$ for each $m_F$, where $F$ and $m_F$ are the quantum numbers of $|F,m_F \rangle$, the lower-angular-momentum eigenstate of the two involved (for the construction of these quantum numbers see Refs.\ \cite{Ledbetter2011PRL107,Theis2013CPL580}).  As demonstrated in Figure \ref{fig:waterFA}c, the $\sim 222$ Hz transition in [$^{13}$C]-formic acid -- corresponding to $F=0 \leftrightarrow 1$ -- splits into two lines ($F$=0, $n$=2) in the bias field.  Spectra of $^{13}$C-methanol ($^{13}$CH$_3$OH) display similar patterns: in zero field the $^{13}$CH$_3$ group gives rise to two observable transitions, one at $^1J_{\mathrm{CH}} \approx 140$ Hz ($F$=0 to $F$=1) and one at $2\times\, ^1J_{\mathrm{CH}} \approx 280$ Hz ($F$=1 to $F$=2).  The perturbing field splits these into a doublet ($F$=0, $n$=2) and a sextet ($F$=1, $n$=6), respectively.  These patterns reveal the quantum numbers involved in each transition, so they may be used to assist assignment and fitting of complicated spectra, or lift ambiguity about the chemical structure of the sample.

At higher fields up to around $10^{-4}$ T the strength of the Zeeman interaction becomes comparable to the J-couplings and the complexity the NMR spectrum for coupled heteronuclear spins can rapidly increase.  The NMR spectra at these fields often contain an even larger number of resonances for the number of distinct spin-spin couplings.  The form of such spectra has been studied theoretically and experimentally measured using both SQUID and vapor cell magnetometers.\cite{Appelt2010PRL81,Bernarding2006JACS128,Bevilacqua2016JMR263,Trahms2010MRI28,Shim2014JMR246}  As an example, $^{13}$C methanol in a field of $5\times 10^{-6}$ T yields a spectrum that contains approximately 10 observable transitions in the frequency range 0-400 Hz, yet the molecule has only one distinct $J_\mathrm{CH}$ coupling.\cite{Shim2014JMR246}

\section{Hardware}

\subsection{Magnetic shielding}

\begin{figure}%
\includegraphics[width=\columnwidth]{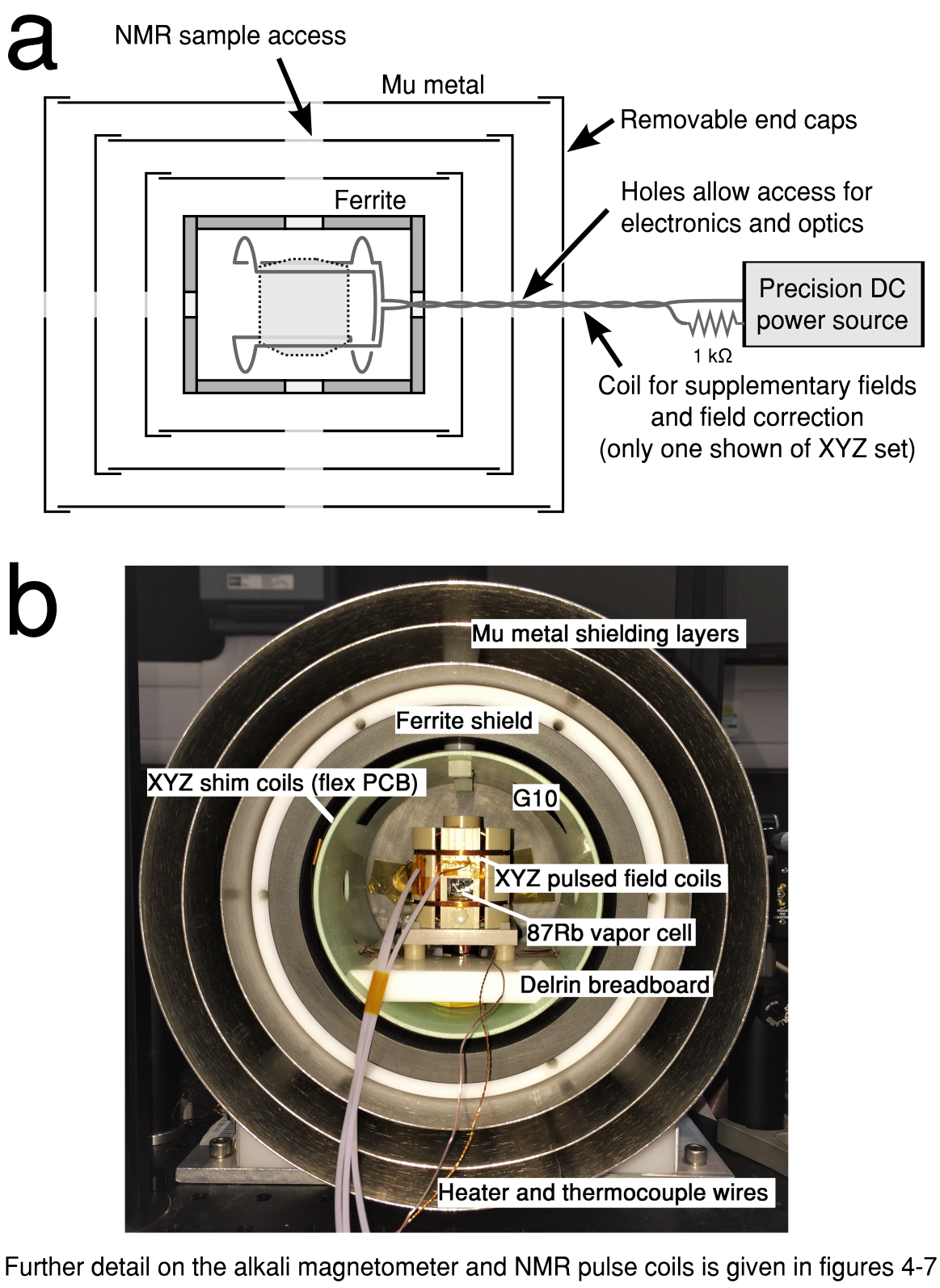}%
\caption{Apparatus for generating the zero/ultra-low magnetic field. In (a) three layers of mu metal attenuate the ambient field of the enclosed volume down to a level of 0.1 nT.  Currents on the order milliamps to microamps can be applied through a set of coils to produce fields along the $x$, $y$ or $z$ axes between 0.1 and 1000 nT within the dotted/shaded region.  The photograph in (b) shows an example of a commercially available shield set (Twinleaf LLC, model MS-1F, 8-inch diameter outer shield), viewed from one of the ends after removal of the end caps.}%
\label{fig:shields}%
\end{figure}

A magnetic field of $\sim 10^{-10}$ T -- below which we accept is a ``zero field'' for liquid-state samples in the experiments concerned -- is established by multiple concentric layers of magnetic shielding as illustrated in Figure \ref{fig:shields}a.  Depending on configuration and size of both the magnetometer and the NMR sample, 3 to 5 layers of mu-metal shielding are used on the exterior.  The photograph in Figure \ref{fig:shields}b shows an example of commercial shields (model MS-1F, Twinleaf LLC) and home-built interior components.  The layers of mu-metal collectively attenuate the ambient magnetic field (e.g.\ of the earth) so the enclosed region is shielded by a factor 10$^5$-10$^6$.\cite{BudkerOpticalMagnetometry}  At this level of shielding, $^1$H Larmor frequencies are below 10$^{-3}$ Hz.  An innermost shield made of ferrite is used to minimize Johnson noise.  Inside the shield there is a set of coils in order to produce fields oriented along the $\pumpaxis$ (solenoid coil, $1.5\times 10^{-4}$ T/A), $\probeaxis$ and $\sensaxis$ axes (saddle coils, $8.0\times 10^{-5}$ T/A) to cancel any remaining field around the cell, or supply bias fields up to around $10^{-6}$ T.  The coils are a set of copper traces printed on Kapton, rolled into a cylinder.  Current is supplied to the coils from low noise, precision sources (KrohnHite model 523).  The internal cylindrical volume measures approximately 150 mm in length and 100 mm in diameter.

Care should be taken to exclude objects with permanent magnetization from proximity to the mu-metal and ferrite layers.  The shielding performance of both materials is degraded when a magnetization is acquired and in the worst case the shield may become magnetically saturated.  Weak magnetization of the shields can be removed by using the following degaussing method:  A thick, insulated wire ($<20$ gauge) rated up to at least 10 A is looped through the shield as many times as possible, 20-30 turns normally being sufficient.  This wire is connected in series with two Variacs and the mains power supply (110 V, 60 Hz).  By adjusting the Variac outputs in an alternating fashion -- changing one by a small amount, then the other, and so on -- the current is increased to approximately 10 A and then returned slowly to zero.  One up-down cycle typically takes 5-10 minutes.  The step size in the degaussing current that results from adjusting a Variac is fairly large by adiabatic demagnetization standards.\cite{Thiel2007RSI78}  However, residual magnetization of the shield is small enough for our purposes, provided at low current the fields are reduced slowly.

The magnetic shields also contain additional access holes of approximately 15 mm diameter to allow entrance of the pump/probe laser beams, the NMR sample and electrical wiring required for the field coils and the vapor cell heater (see Section \ref{sec:detectionsetup}).  These holes do not appear to significantly affect the shielding quality in the central working region.

\subsection{Polarization}

\subsubsection{\emph{Ex situ}: Zeeman polarization}
\label{sec:zeemanpolarization}
The NMR experiment must begin with polarized nuclear spins, i.e.\ populations of the nuclear spin states must not be equal.  The most general way to initialize spin polarization is to couple the nuclei to a magnetic field ($\bm{B}_\mathrm{pol}$, via the Zeeman interaction) and allow the populations of the spin eigenstates to reach thermal equilibrium.  For field strengths $10^{-3}$ T$<|B_\mathrm{pol}|<10^2$ T and temperatures $T>1$ K the thermal polarization of a single spin-1/2 can be approximated by 
\begin{equation}
p = \gamma |\bm{B}_\mathrm{pol}| / 2 \pi k_B T,
\end{equation}
where $k_B = 2.084 \times 10^{10}$ Hz K$^{-1}$ is Boltzmann's constant and the orientation is parallel to $\bm{B}_\mathrm{pol}$.  The normalized spin-1/2 density operator under these conditions is given by $\rho_\mathrm{eq} \approx (1 + p \bm{I}\cdot\bm{B}_\mathrm{pol}/|\bm{B}_\mathrm{pol}|)/2$ ($\mathrm{tr}(\rho)=1$).  Thus, $^1$H nuclei at room-temperature in a field of 2 T will acquire an equilibrium polarization of around 14 ppm.  This leaves considerable room for improvement since, by definition, the polarization can be as large as $|p|=1$.  Nevertheless, for many of the basic experiments described in this paper the method is sufficient.  The level of polarization is also highly consistent, as long as time is left for the thermal equilibrium to be reached, thus suitable for multi-dimensional NMR and experiments involving phase cycles or other types of signal addition.

\begin{figure}[t]%
\includegraphics[width=\columnwidth]{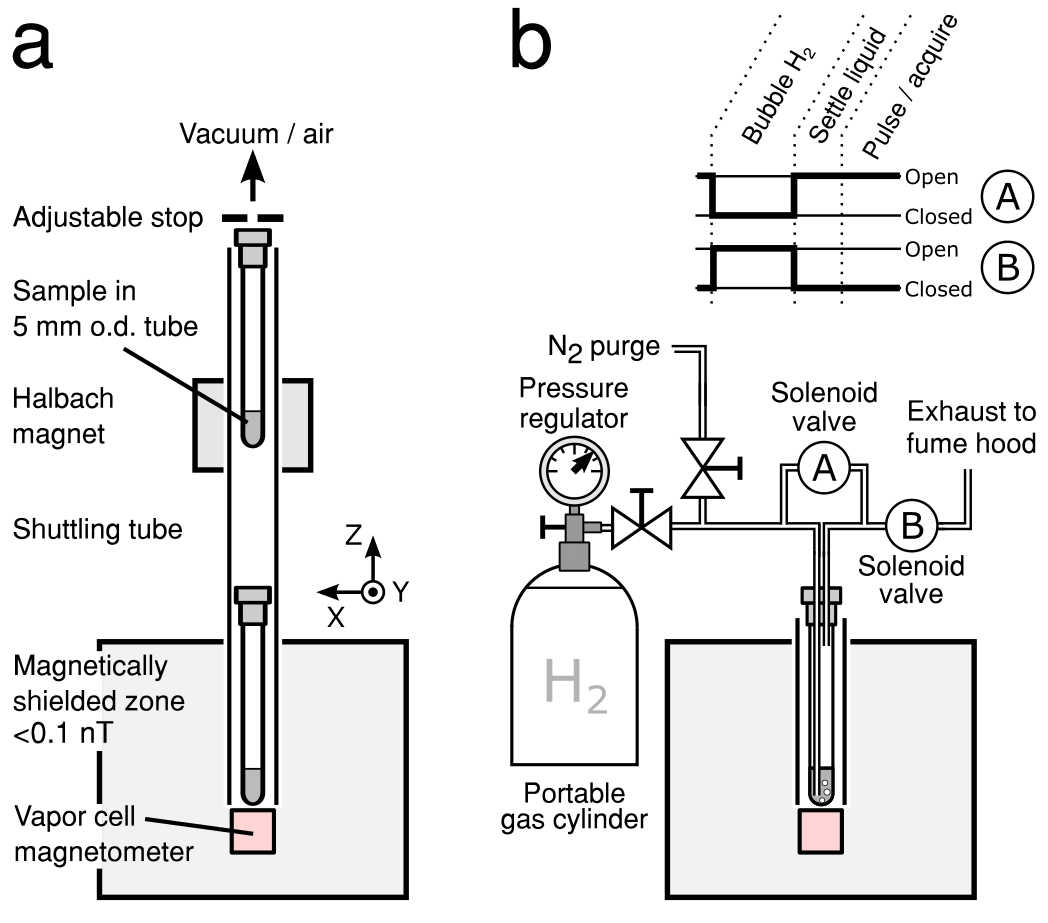}%
\caption{Schemes for nuclear spin polarization of a substrate prior to a ZULF NMR experiment: (a) thermal prepolarization in a permanent magnet, followed by shuttling; (b) polarization via interaction with para-hydrogen.}%
\label{fig:polarization}
\end{figure}

As explained earlier strong magnetic fields (above $10^{-1}$ T) should not be applied within the magnetic shields, so the procedure for polarizing the nuclear spins requires an \emph{ex situ} approach, unless a hyperpolarization scheme is used that works at ultralow field, as exemplified in the next section.  The sample is polarized in a magnet located a safe distance away from the shields and is then shuttled inside.  In our setup a 2 T Halbach array magnet is placed on a shelf above the shields at a distance of 10-20 cm.  The sample is placed in a standard 5 mm outer diameter (o.d.) NMR tube and is shuttled between the magnet and the center of the shield inside a fiberglass tube whose inner diameter (i.d.) is slightly larger than 5 mm, and whose o.d.\ easily fits through the holes in the shielding material.  This arrangement is illustrated in Figure \ref{fig:polarization}a.  The sample can be raised for subsequent re-polarization by applying suction to the top of the tube.  Regulation of air pressure at the top and base of the fiberglass tube controls rates of ascent and descent.  Fluidic transport may be used as alternative: two containers may be positioned at fixed locations in the permanent magnet and next to the vapor cell magnetometer and the liquid sample pumped between these.\cite{Savukov2005PRL}   Other setups may be engineered but have not been tested.

The timescale of the shuttling operation, specifically the speed of switching between different field regimes, is an important consideration in a ZULF NMR experiment.  The sample should be shuttled much faster than the relaxation time of the spins so as to preserve as much of the initial polarization as possible.  This can be problematic for molecules containing nuclei with nonzero quadrupole moments (including $^2$H, $^{14}$N and all halogens except fluorine) since scalar relaxation of the second kind\cite{Abragambook1961} is a significant mechanism at low field causing near-complete loss of the spin polarization within a few tens of milliseconds.  A second consideration is the change in direction and magnitude of background magnetic fields, including the fringe field of the polarizing magnet and the field due to the earth.  The transport can be performed adiabatically (slow relative to the spin dynamics) or sudden (fast relative to spin dynamics).  The difference is that an adiabatic transfer produces polarization in the ZULF eigenstates of the spin system, whereas a sudden transfer leaves the initial density matrix unaffected and may lead to detectable coherences as soon as the sample arrives in the ZULF detection region.  

These differences can be seen in how density operator is influenced by the rate at which the sample is transported through the various magnetic field gradients.  In particular, if the rate of frequency change $|d\nu/dt|$ is similar to the nuclear precession frequency $\nu_0$ and/or the spin-spin couplings, then ill-controlled spin dynamics result.  Therefore, the transport either has to be much faster (nonadiabatic, $|d\nu/dt|\gg |\nu_0|,|J|$) or much slower (adiabatic, $|d\nu/dt|\ll |\nu_0|,|J|$) than the spin dynamics.  

If the transit between the polarizing field and ZULF is rapid, the high-field thermal polarization is ``instantly'' brought into ZULF and ignoring relaxation is essentially unchanged.   For our example of [$^{13}$C]-formic acid, the density operator would be approximately
\begin{eqnarray}
\rho &=& (1 + p_H I_{z}^{\mathrm{H}} + p_C I_{z}^{\mathrm{C}})/4,
\end{eqnarray}
where the polarizations are $p_H = (\gamma_H B /2\pi) / (k_B T)$ and $p_C = (\gamma_C B /2\pi) / (k_B T)$.  Even for this simple system, $\rho_(0)$ does not commute with the zero-field spin Hamiltonian (the operator for scalar coupling, $\bm{I}^{\mathrm{H}}\cdot \bm{I}^{\mathrm{C}}$) due to the component $(I_{z}^{\mathrm{H}} - I_{z}^{\mathrm{C}})$ and therefore sudden switching generates spin coherences between the ZULF eigenstates that result in a time-dependent magnetization.  Sudden switching is achievable by very rapid sample movement where the transport is completed within $\sim 10$ ms.\cite{Biancalana2014RSI85} Alternatively, the sample can be shuttled slower through a ``guide field'' on the order of $10^{-5}$-$10^{-4}$ T along the shuttling path through a solenoid coil wound along the length of the fiberglass shuttling tube.  The field maintains a Zeeman eigenbasis until the sample is stationary next to the magnetometer, inside the shields.  At this point the solenoid field is rapidly quenched ($< 10$ ms), to convert suddenly between the Zeeman (high field) and ZULF eigenbases.

At the other extreme, if the sample if shuttled slowly into the low field  ($\nu^{-1}|d\nu/dt|\ll \nu$) the transport is adiabatic.  It is desirable to transport the polarized spins adiabatically through the fringe fields (meaning the field orientation changes slowly relative to the speed of nuclear precession) such that the spins remain oriented along the axis of total magnetic field, which is not necessarily parallel to $\bm{B}_\mathrm{pol}$.   The populations of the high-field eigenstates are then smoothly transferred to the ZULF eigenstates, resulting in a new density operator:\cite{Emondts2014PRL112}  
\begin{eqnarray}
\rho &\rightarrow& \Bigl( 1 + \frac{p_H + p_C}{2} (I_{z}^{\mathrm{H}} + I_{z}^{\mathrm{C}})  \nonumber \\ 
&& \qquad + \frac{p_H - p_C}{2} \bm{I}^{\mathrm{H}}\cdot \bm{I}^{\mathrm{C}} \Bigr)/4.
\label{eq:rhoadiabatic}
\end{eqnarray}
The adiabatic method does not immediately produce an NMR signal at zero field because the density matrix operator commutes with the spin Hamiltonian.  Subsequent manipulations, such as pulsed fields are required, which are described in section \ref{sec:encoding}.

\subsubsection{\emph{In situ}: parahydrogen-induced polarization (PHIP)}
A different method of polarizing the sample is to expose it to para-enriched dihydrogen (H$_2$), which contain up to unity polarization in the nuclear singlet state of the proton pair. Polarization from H$_2$ may be transferred to the target molecule via spin-spin couplings.  The enhancement can be several orders of magnitude beyond the Zeeman polarization obtained by magnets.

There are numerous ways to create these couplings. One is a hydrogenation reaction, such as addition of polarized H$_2$ molecules at double or triple bonds in the target molecule.\cite{Bowers1986PRL57,Bowers1987JACS109}  A non-hydrogenative way involves reversible complexation at a metal ion, where both H$_2$ and the substrate bind as ligands and spin coupling arise due to the presence of a common set of molecular orbitals.\cite{Adams2009Science323}  Usually, a reversible reaction is desired, in order to polarize the spins without altering the chemical properties or structure of the analyte.  In both types of reaction, PHIP enhancements depend strongly on the physical conditions, the chemical nature of the analyte, the kinetics of the spin order transfer and the relaxation times.  For molecules incorporating $^{15}$N or $^{31}$P lone pairs,\cite{Colell2017JPCC} the non-hydrogenative route is efficient in the regime of 1 to 10 $\mu$T where the J couplings to para-hydrogen are similar in magnitude to the Zeeman interaction, leading directly to polarization of the heteronuclei.\cite{Theis2015JACS137,Zhivonitko2015CCommum51}  

The appeal of parahydrogen induced zero-field NMR experiments is that sample shuttling is avoided establishing ``NMR without magnets'' as it was coined in the popular press.\cite{Theis2011Nature473} A scheme is shown in Figure \ref{fig:polarization}.  The substrate and catalyst are dissolved in solution and contained in a 5 mm o.d.\ NMR tube.  Para-enriched H$_2$ is bubbled into this solution through a 1/32 inch o.d.\ Teflon tube, where the flow is controlled by the two solenoid valves ``A'' (normally open) and ``B'' (normally closed) triggered by the data acquisition system.  Hydrogen is bubbled for a timed interval, during which the reaction occurs, then the liquid is allowed to settle for 100 ms before finally the NMR experiment is started.  

With these methods, $^{13}$C NMR signals can be detected at the low natural abundance of $^{13}$C ($1.1\%$)\cite{Theis2011NatPhys7,Butler2013JCP138} and also $^{15}$N spectra can be obtained at the even lower natural abundance of $0.36 \%$,\cite{Theis2012JACS134} enabling the application of ZULF NMR to the spectroscopy of organic molecules.  Detection of the NMR signals in ZULF may play a central role in understanding the details of the hyperpolarization mechanism in low field because the hyperpolarization is detected directly at the location where it is created.

\subsection{Encoding}
\label{sec:encoding}

\subsubsection{Pulsed field excitation}
\begin{figure}%
\includegraphics[width=\columnwidth]{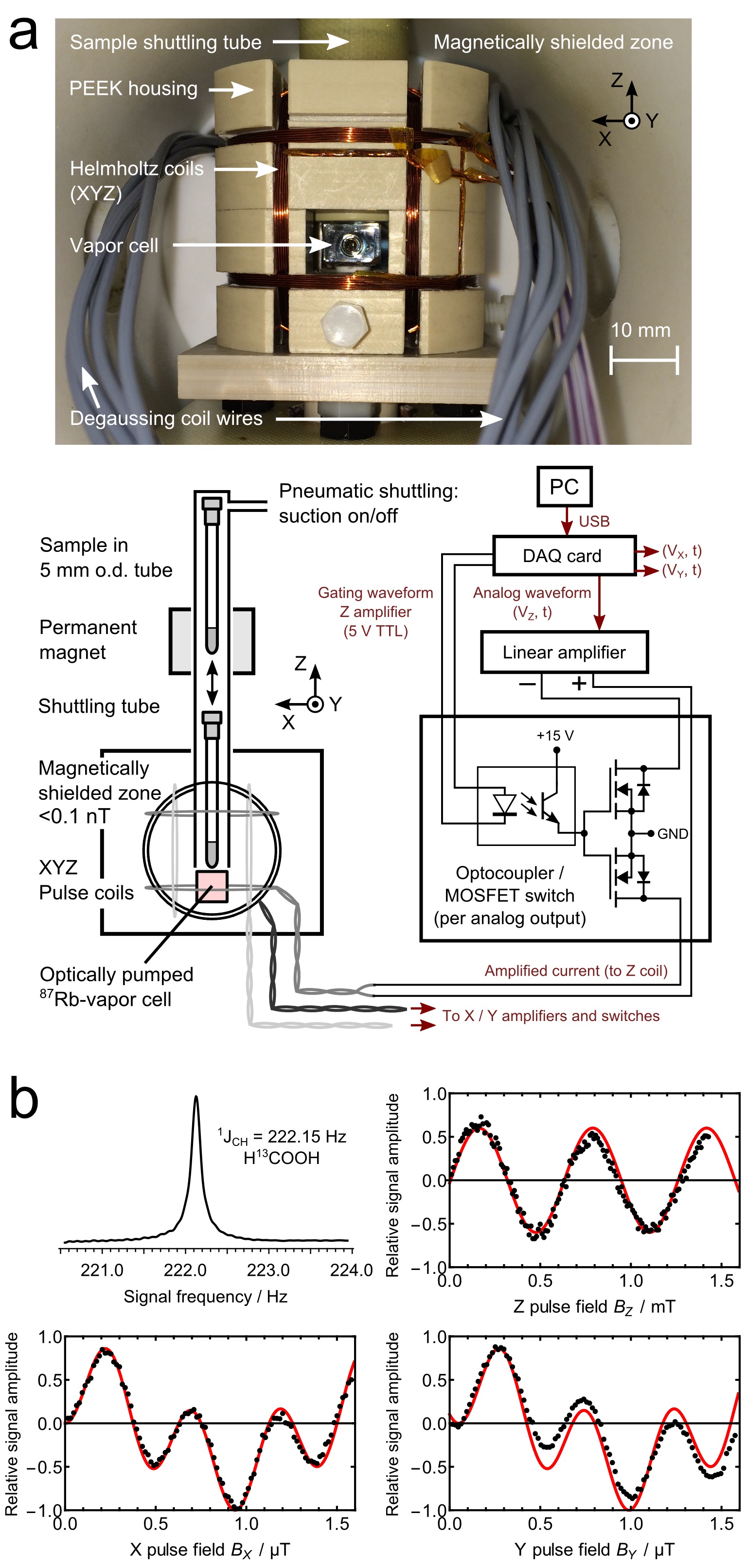}%
\caption{Illustration of the pulsing coil setup in the ZULF NMR spectrometer: (a) photograph and schematic showing the arrangement of the coils around the vapor cell and control interface; (b) zero-field NMR signal amplitude of [$^{13}$C]-formic acid versus amplitude of a DC excitation pulse, following prepolarization and adiabatic shuttling into zero field.  Pulses were applied via the $x$, $y$ or $z$ axis coils, with a duration a 50 $\mu$s.}%
\label{fig:ZFNMRinstrument}%
\end{figure}

For NMR pulse sequences we use three orthogonal sets of Helmholtz coils (28 AWG enameled Cu wire, 3 cm diameter, 6 turns, 230 $\mu$T/A, 1.0 $\Omega$) inside the shields, as shown in Figure \ref{fig:ZFNMRinstrument}a.  The coils are wound on grooves that are machined in a polyether-ether-ketone (PEEK) block and are positioned such that the fields generated are centered on the sample when just above the magnetometer and are oriented along the laboratory $x, y$ (horizontal) and $z$ (vertical) directions as indicated.  To apply pulses, the desired waveform is written as a list of time/voltage coordinates $(t, V_x, V_y, V_z)$ in a file on a computer.  A microcontroller and digital-analog converter (e.g. National Instruments USB 6229) converts these data into a three-channel analog output voltage (16-bit, $-5$ V to +5 V).  Each output channel is fed into a low-distortion controlled-current amplifier (AE Techron LVC2016, maximum output 10 A at 1.0 $\Omega$ load), whose output is connected to one of the coils.  In series with each coil are 5V-logic bipolar field-effect transistors whose switching time is 1-10 $\mu$s (lower-right part of Figure \ref{fig:ZFNMRinstrument}a).  To minimize magnetic field noise introduced by the coils to the region inside the shields -- in particular the $\sim$ 1 mA noise of the current amplifiers -- these switches permit the coils to only form a closed circuit when pulses are being applied.

In many NMR experiments it is necessary to perform spin-selective reorientations and ZULF is no exception.  Constant-amplitude direct-current pulse waveforms can suffice for this job as follows.  If we ignore the effects of spin-spin couplings, each spin precesses about the net field $\bm{B}_\mathrm{pulse}$ with an angular velocity $\gamma_i \bm{B}_\mathrm{pulse}$ that depends on the gyromagnetic ratio, thus allowing a change in relative orientation.  Ignoring the spin coupling terms is a good approximation provided that the pulses satisfy $\tau_\mathrm{pulse} |J| \ll 1$ and $|\gamma B|/2\pi\gg|J|$, which is usually the case for $|\bm{B}| > 100$ $\mu$T and pulse flip angles on the order of a few radians.  

In [$^{13}$C]-formic acid, a DC field along the $x$ axis $\bm{B}_\mathrm{pulse} = \{ B_x,0,0\}$ for a duration $\tau_\mathrm{pulse} = 4\pi/|\gamma_\mathrm{H} B_x|$ would result in rotations of $^1$H and $^{13}$C spin polarization about $x$ through angles $4\pi$ and $4\pi |\gamma_\mathrm{C}/\gamma_\mathrm{H}| \approx \pi$, respectively.  This pulse allows an observable signal to be excited from the postcursor state after adiabatic shuttling, given in Equation \ref{eq:rhoadiabatic}.  We can see that a 180-degree rotation of either spin species converts $(I_z^\mathrm{H}+I_z^\mathrm{C})$ into $\pm(I_z^\mathrm{H}-I_z^\mathrm{C})$, which as explained in the introduction is nonstationary under the zero-field Hamiltonian.

Figure \ref{fig:ZFNMRinstrument}b shows the amplitude of the zero-field NMR signal for [$^{13}$C]-formic acid after a 50 $\mu$s pulse, versus the pulse-field amplitude along the $x$, $y$ and $z$ coil axes.  The change in spin orientation changes the coefficient of the spin order $(I_z^\mathrm{H}-I_z^\mathrm{C})$ and thus the signal with pulse length $\tau_\mathrm{pulse}$ as $\sin[(\gamma_\mathrm{H} + \gamma_\mathrm{C}) B_x \tau_\mathrm{pulse}/2]\sin[(\gamma_\mathrm{H} - \gamma_\mathrm{C}) B_x \tau_\mathrm{pulse}/2]$ which is plotted as the solid curve.  The signal after an $x$- and $y$-field amplitude is ideally the same, since the amplitude of $(I_z^\mathrm{H}-I_z^\mathrm{C})$ that results from each rotation does not depend on the direction of the field in the $xy$ plane.  

The dependence on field amplitude $B_z$ along the $z$ axis follows the function $\sin((\gamma_\mathrm{H} - \gamma_\mathrm{C}) B_z \tau_\mathrm{pulse})$, which is different because it is the product term $\bm{I}^{\mathrm{H}}\cdot \bm{I}^{\mathrm{C}}$ (c.f.\ Equation \ref{eq:rhoadiabatic}) that generates the observable, rather than the orientation parallel to $z$.  Overall, these curves provide an accurate calibration of magnetic field vs applied current to the coils, which is needed in order to implement sequences of many pulses where precise nuclear spin reorientation is demanded, including spin-echo pulse trains\cite{Lee1987JMR75} and spin decoupling.\cite{Llor1991PRL67,Sjolander2017JPCLXX}  Pulse trains also offer the option to measure the zero-frequency component of the NMR signal by ``lock-in detection'': the nuclear magnetization is periodically inverted using 180$^\circ$ pulses to produce a signal looking like a square wave, which is then separable from DC offsets.\cite{Ganssle2014ANIE53}

The pulsing setup outlined in Figure \ref{fig:ZFNMRinstrument}a allows the amplitude of applied fields to be varied smoothly, enabling one to implement conventional ``high-field'' NMR methodology in which resonant AC fields at the Larmor frequency are used to obtain spin species selectivity: a DC pulse ($|\bm{B}_\mathrm{pulse}| < 2$ mT, for several seconds without noticeable heating effects) is applied along one axis, and at the same time a much weaker AC pulse ($|\bm{B}_\mathrm{pulse}|_\mathrm{max} < 0.02$ mT) is applied with the Larmor frequency of the selected spins, up to several tens of kHz.\cite{Tayler2016JMRXX}  A larger number of coil windings would allow DC fields of several tens of millitesla to be reached, where chemical shifts are large enough to be resolved and exploited.  For example at $10$ mT, a 100 ppm chemical shift of $^{13}$C corresponds to a frequency difference of 10 Hz.

\subsubsection{Extremely-low-frequency pulses}
Strong DC pulses of the type described above and illustrated in Figure \ref{fig:ZFNMRinstrument} are broadband with respect to the zero-field spectrum.  The reason for this is that the Larmor precession frequencies of the sample spins in the applied fields are much greater than the eigenfrequencies of the intramolecular Hamiltonian ($|(\gamma_j \pm \gamma_k )\bm{B}_\mathrm{pulse}| \gg |2\pi J_{jk}|$) that is the source of the ultra-low-field signal.  Therefore, such pulses affect all spins in the spectrum.
 
If one wishes to exert control over individual transitions, one should use low-amplitude modulated pulsed fields that are much weaker than the internal Hamiltonian and resonant with a transition of choice.\cite{Sjolander2016JPCAXX}  It can be shown that under these conditions the spin order is driven selectively between spin eigenstates connected by the resonant field.  The rate of these transitions are calculable given the total and projection angular momentum quantum numbers of the states involved and the strength of the resonant field.

\subsection{Detection}

\subsubsection{Principles of optical magnetometry}
\label{sec:detectionintro}
To probe the spin dynamics we detect the magnetic field emanating from the NMR sample with an atomic magnetometer.  We exploit an effect known as \emph{magneto-optical rotation}.  This effect involves the passage of linearly polarized light through a medium and its subsequent rotation due to the difference of refractive indices $n_{\pm}$ for the two circularly polarized components ($\sigma^\pm$) of the light.  For a path length $l$ and light frequency $\nu$, the rotation angle is given by  
\begin{eqnarray}
\varphi = (n_{+} - n_{-}) \pi \nu l/c.
\end{eqnarray}
If the difference in refractive indices $(n_{+} - n_{-})$ depends on the magnetic field at the material, the circular birefringence allows the magnetic field to be determined.   

In our case, an atomic vapor of an alkali metal is probed near the D1 atomic resonance ($^2$S$_{1/2}$ to $^2$P$_{1/2}$ transition) and the refractive index is bestowed by absorption of light, characterized with a sharp frequency dependence near resonance.  We may consider the D1 transition without hyperfine effects and so assume a four-level system comprising the excited-state ($^2$P$_{1/2}$) and ground-state ($^2$S$_{1/2}$) doublets, each with projections of the atomic spin $m_J=-1/2$ and $m_J=+1/2$.  Light that is circularly polarized along the quantization axis may induce transitions $\Delta m_J = \pm1$, where the sign indicates interaction with $\sigma^+$ or $\sigma^-$ light.  

A difference $(n_{+} - n_{-})$ may be encoded with a longitudinal magnetic field $B$ parallel to the light beam because the absorption maxima of the two transitions $\Delta m_J = \pm1$ shift in opposite directions: $\nu_0^\pm = \nu_0 \pm g \mu_B B$.  Here $g$ is the gyromagnetic ratio of the atom (ground state Land\'{e} factor) and $\mu_B$ is the Bohr magneton.  The dependence of the real refractive indices $n_\pm$ on $B$ is related to the the imaginary part of the (Voigt) absorption lineshape, $\mathcal{V}(\nu-\nu_0^\pm)$, via the Kramers-Kronig relations.  Taking into account the populations $\rho_\pm$ for the ground states $m_J = \pm1/2$, the difference can be expressed as
\begin{eqnarray}
(n_+ - n_-) &\propto& \rho_-\mathrm{Im}[\mathcal{V}(\nu - g \mu_B B - \nu_0)] - \nonumber \\&&\qquad \rho_+\mathrm{Im}[\mathcal{V}(\nu + g \mu_B B - \nu_0)]
\end{eqnarray}
so at any frequency $\nu$ within the resonance line the refractive indices are unequal.  This phenomenon for the case of resonant absorption in atomic vapors is named the Macaluso-Corbino effect.\cite{Macaluso18981,Macaluso18982,Zeeman1897Nature55}

\begin{figure}%
	\includegraphics[width=0.9\columnwidth]{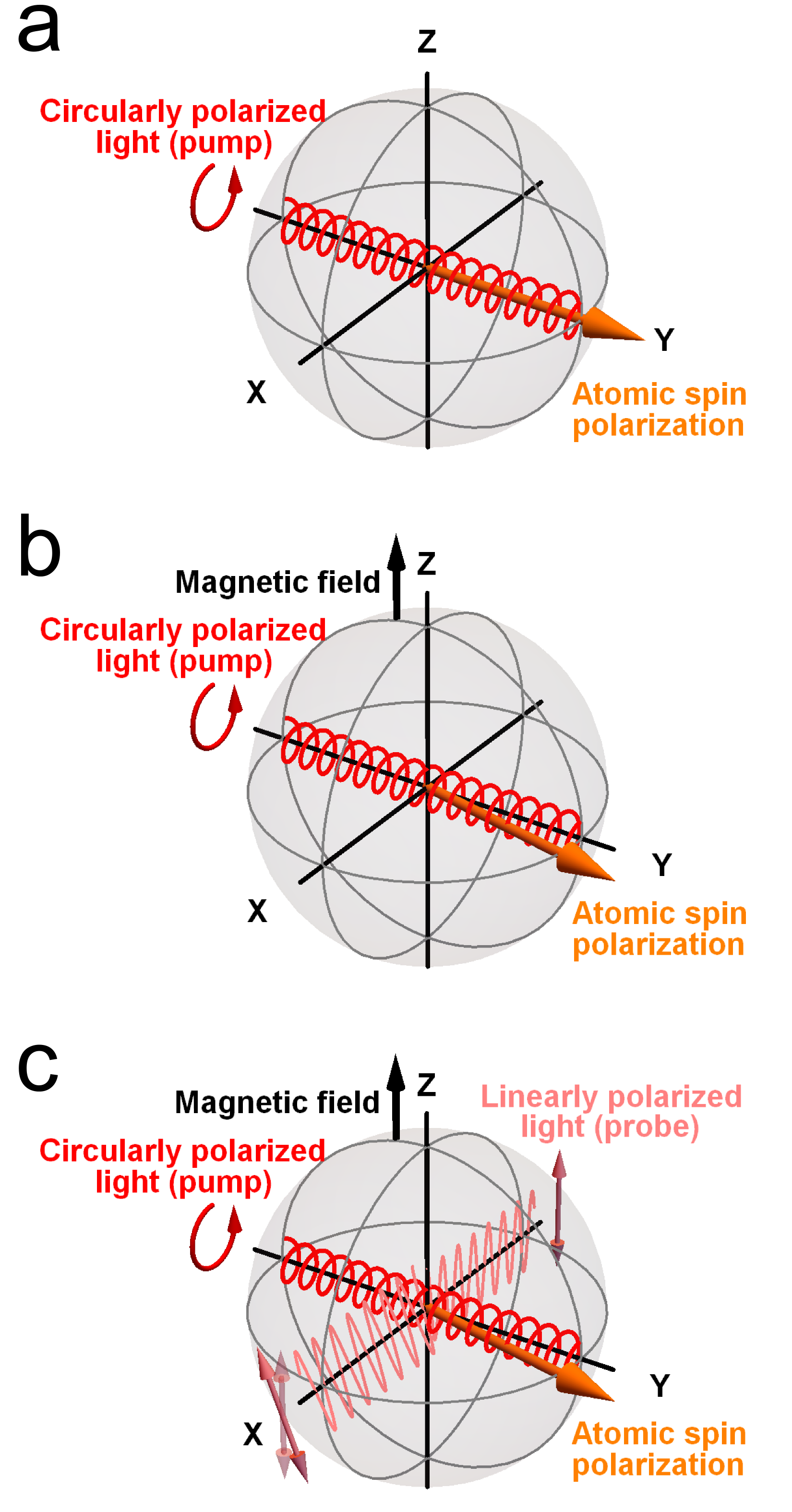}%
	\caption{Scheme for optical pump-probe magnetometry.  The orange-colored arrow represents the average spin polarization $\langle S \rangle$ of alkali atoms in the magnetometer: (a) incident circularly polarized pump light induces polarization along the $\pumpaxis$ axis; (b) an off-axis component of steady-state polarization arises in a finite magnetic field $B_\sensaxis$ due to spin precession: $\langle S_\probeaxis \rangle \propto B_\sensaxis$.  In (c) plane-polarized probe light beam passes through the vapor along the $\probeaxis$ axis.  The plane of polarization is rotated by an angle $\varphi \propto \langle S_\probeaxis \rangle$, allowing $B_\sensaxis$ to be measured.}%
	\label{fig:odmroverview3}
\end{figure}

Our magnetometer involves a strongly amplified version of the Macaluso-Corbino effect, which exploits nonlinear optical properties of the alkali atom vapor.  As well as shifting the frequencies of the $\Delta m_J = \pm1$ transitions, the absorption coefficients are strongly debalanced by optical pumping.  The optical pumping stems from interaction of the atoms with circularly polarized light, which induces redistribution of atomic population among the Zeeman sublevels.  As the sublevels correspond to specific projections of spin onto the quantization axis, this leads to spin polarization of the medium.  The resulting magneto-optical effect is summarized in Figure \ref{fig:odmroverview3} and explained below.\cite{Happer1967PR163,Happer1972RMP44}  

If the pump light propagates with circular polarization $s$ along the $\pumpaxis$ axis (unit vector $\bm{e_\pumpaxis}$) of the instrument, the atomic polarization $\bm{S}$, determining the optical anisotropy, is determined by the Bloch equation
\begin{eqnarray}
d \bm{S} /dt &=& [g (\mu_B/\hbar) \bm{B} \times \bm{S} \nonumber \\ && \qquad + R_{\mathrm{OP}} (\bm{e_\pumpaxis}/2 - \bm{S}) - R_{\mathrm{rel}} \bm{S}]/q, \label{eqn:blochalkali}
\end{eqnarray}
where $q$ denotes the nuclear slowing-down factor, $R_{\mathrm{OP}}$ is the rate of optical polarization and $R_{\mathrm{rel}}$ is the rate of polarization decay in the ground state.\cite{BudkerOpticalMagnetometry}  For the case of zero field, $\bm{B}=\bm{0}$, a steady-state atomic polarization ($\mathrm{d}\bm{S}/\mathrm{d}t=\bm{0}$) is given by $\langle S_\pumpaxis \rangle = S_0 = (+1/2)R_{\mathrm{OP}}/(R_{\mathrm{OP}}+R_{\mathrm{rel}})$ and $\langle S_\probeaxis \rangle = \langle S_\sensaxis \rangle = 0$ (Figure \ref{fig:odmroverview3}a).  

For a nonzero value of $\bm{B}$, atomic spin precession causes a component of the steady-state polarization that is perpendicular to both $\bm{B}$ and the pump beam axis.  For a field $\bm{B} = \bm{e_\sensaxis} B_\sensaxis$, the steady-state solution of Equation \ref{eqn:blochalkali} yields a finite component $\langle S_\probeaxis \rangle = S_0 (g \mu_B / \hbar) B_\sensaxis$, such as that illustrated in Figure \ref{fig:odmroverview3}b.  

To probe the polarization component $\langle S_\probeaxis\rangle$, the linearly polarized probe beam is aligned with the $\probeaxis$ axis (Figure \ref{fig:odmroverview3}c).  The component $\langle S_\probeaxis \rangle$ corresponds to a ground-state population difference across the $m_J = \pm 1/2$ ground states.  In such a configuration the polarization of the probe is a direct measure of the magnetic field.  The rotation is enhanced by many orders of magnitude compared to the linear Macaluso-Corbino effect.  It is hence called nonlinear magneto-optical rotation.

In general, Equation \ref{eqn:blochalkali} can be solved to give the rotation angle $\varphi$ as a function of the magnetic field in any direction.  The probe beam polarization, giving the magnetometer signal, is sensitive to fields along all axes $x$, $y$ and $z$ according to the function
\begin{eqnarray}
\varphi \propto S_\probeaxis = S_0 \frac{(\Delta B) B_\sensaxis + B_\probeaxis B_\pumpaxis}{\Delta B^2 + B_x^2 + B_y^2 + B_z^2},\label{eq:shimxyz}
\end{eqnarray}
that is a dispersive Lorentzian with a linewidth $\Delta B = (g \mu_0/\hbar)/(R_\mathrm{op}+R_\mathrm{rel}) $ when $B_\pumpaxis=B_\probeaxis=0$.\cite{BudkerOpticalMagnetometry}  However, the magnetometer is most sensitive to  $B_\sensaxis$ when operated in the region $|B_\pumpaxis|\ll \Delta B, |B_\probeaxis|\ll \Delta B$, where the slope $\mathrm{d}\varphi/\mathrm{d}B_\sensaxis$ is largest ($S_\probeaxis \approx S_0 B_\sensaxis / \Delta B$).  
The magnetic field generated by an NMR sample is generally not larger than 1 pT.  This lies well within the central part of the dispersion curve ($|B_z| \ll |\Delta B|$) resulting in a magneto-optical rotation where the angle $\varphi$ is linearly proportional to the NMR signal.  This justifies our choice of observable in Section \ref{sec:NMRspindynamics}.  

The magneto-optical effect in this regime can be characterized by a Verdet coefficient ($V$), defined as the rotation angle $\varphi$ at a given wavelength per unit path length $l$ and applied field $B_\sensaxis$: $\varphi = V B_\sensaxis l$.  For Faraday rotation in most solid and liquid materials $V$ does not usually exceed 1000 rad / T$\cdot$m.  For sparse, optically pumped alkali vapors in the SERF regime the Verdet coefficient can be as large as $10^{12}$ rad / T$\cdot$m.  Therefore, on a per-atom basis, the atomic magnetometer is more sensitive by around 11-12 orders of magnitude.  The fundamental sensitivity of the atomic magnetometer is limited by the ability to determine the atomic spin projection due to the Heisenberg uncertainty principle.\cite{Budker2002RMP74}  For a measurement time $T$, the uncertainty in $B$ is given by the expression 
\begin{eqnarray}
\delta B \sim \frac{h}{2\pi g \mu_B} \sqrt{\frac{R_\mathrm{rel}}{N T}},\label{eqn:deltaB}
\end{eqnarray}
where $N$ is the number of atoms involved.  In the literature, $\delta B$ is frequently given per unit bandwidth in units of T$/\mathrm{Hz}^{1/2}$.  Typically, although not universally, a bandwidth of 1 Hz corresponds to a measurement time of $T = 0.5$ s.

Although Equation \ref{eqn:deltaB} implies that large numbers of alkali atoms are favored, there are some caveats.  A dense atomic vapor incurs high rate of atom-atom spin-exchange collisions.  This corresponds to an increase in $R_\mathrm{rel}$.  As the two hyperfine ground states of alkali atoms are characterized with nearly opposite Land\'{e} factors, (they precess in opposite directions in a magnetic field), the spin-exchange-induced transitions between the hyperfine states introduces a relaxation that for a broad range of temperatures (70-170$^\circ$C) prevents from increasing the magnetometric sensitivity.  However, this changes at even higher temperatures and vapor densities.  The SERF regime at ultralow field is characterized by a slow precession of the atomic spins relative to the rate of atom-atom collisions.  Here the hyperfine state is rapidly (and randomly) switched and averaged, such that spin-exchange collisions no longer act as a significant relaxation mechanism.\cite{Happer1977PRA16,Happer1973PRL31}   Relaxation in the SERF regime is consequently dominated by spin-destruction collisions, whose cross section is between 3 and 5 orders of magnitude smaller than for spin exchange, allowing sensitivities that approach down to the spin-projection limit below 1 fT/Hz$^{1/2}$.\cite{Savukov2005PRA71,Allred2002PRL89,Ledbetter2008PRA77,Kominis2003Nature422,Savukov2005PRL,Ledbetter2009JMR199}

Finally, as the NMR signals of interest correspond to time-dependent fields, an important parameter is magnetometer bandwidth.  The response of the magnetometer to a change in the magnitude of $B_\sensaxis$ occurs with the time constant $1/T_2 = (R_\mathrm{op} + R_\mathrm{rel}/q)$ of the alkali atom spins, thus the response to oscillating field versus frequency is a Lorentzian profile with half width $1/(2\pi T_2)$ at half maximum (i.e.\ Fourier transform of the response curve).\cite{BudkerOpticalMagnetometry}  We see that an increased magnetometer bandwidth then comes at the expense of a reduction in sensitivity, since the two quantities scale oppositely in $T_2$.  For our instrument, an acceptable sensitivity is 20-30 fT/Hz$^{1/2}$, which we find is achievable over the 0-500 Hz frequency range.

\subsubsection{Magnetometer setup}
\label{sec:detectionsetup}
\begin{figure}%
	\includegraphics[width=\columnwidth]{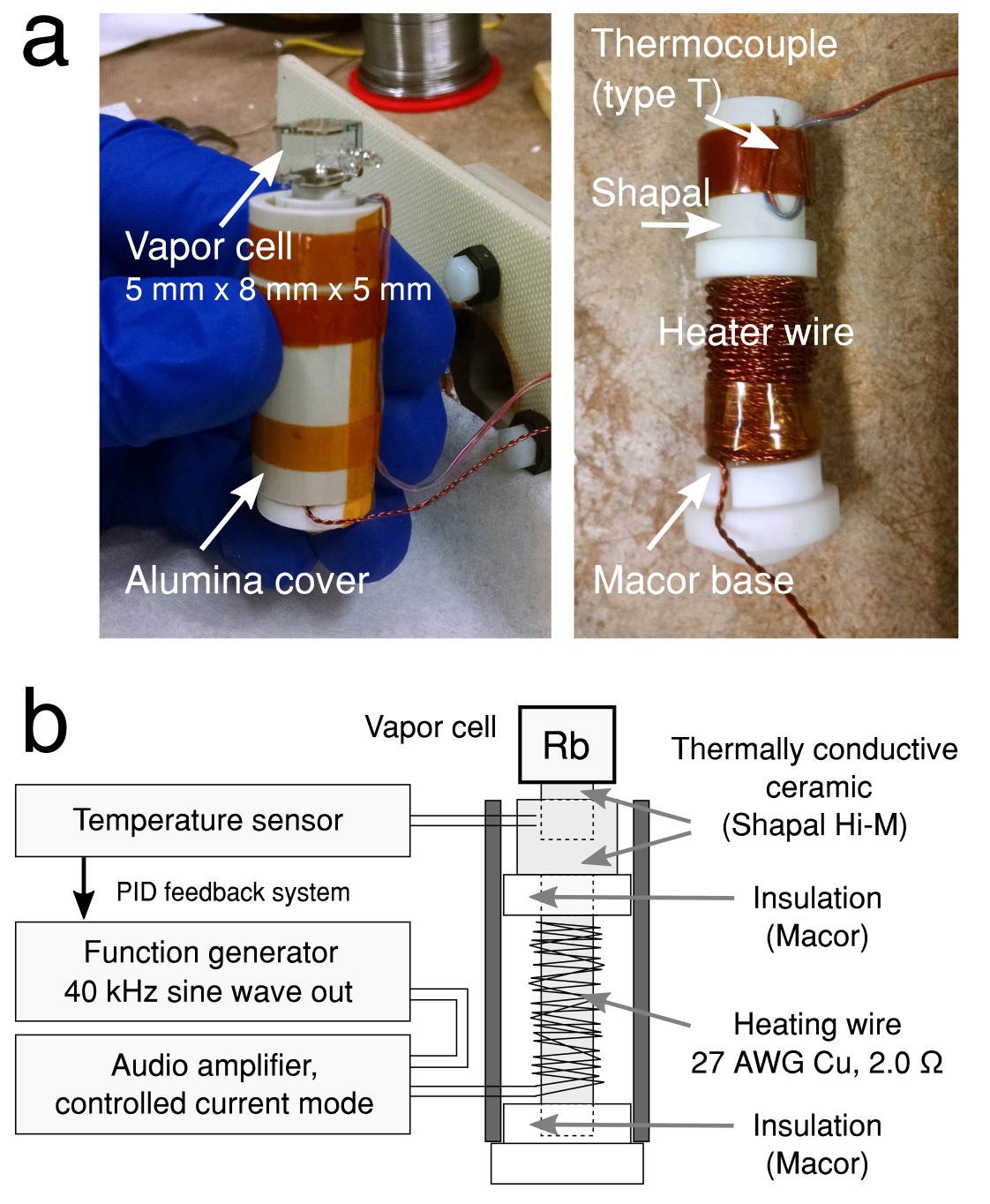}%
	\caption{Details of the heating element and electrical circuit to heat the rubidium vapor cell to its operating temperature.}%
	\label{fig:heater}
\end{figure}

In our setup, rubidium-87 is used as the atomic medium and is confined to a cuboidal cell made from borosilicate glass, inner dimensions measuring $5\times5\times8$ mm$^3$ (2-5 mg $^{87}$Rb of isotopic purity $98\%$, N$_2$ buffer gas at 700 torr = 93 kPa, Twinleaf LLC, see top-left part of Figure \ref{fig:heater}a). This cell is uncoated on its walls and is heated to temperatures of 170-190 $^\circ$C to achieve a sufficient vapor density of the alkali atoms and rate of spin-exchange collisions for maximum sensitivity in the SERF regime.  The nitrogen acts as a buffer gas to accelerate optical pumping of the alkali atoms and maximize the overall polarization by avoidance of radiation trapping.\cite{MolischOehrybook}

The operating temperature of the Rb cell is reached by resistive-inductive heating through a 5-meter length of double-twisted 27-gauge enameled copper wire (of resistance around 2 $\Omega$), which is wrapped approximately 50 turns around a thermally conductive cylindrical spool made from aluminum nitride (Shapal Hi-M, Precision Ceramics, room-temperature thermal conductivity 92 W/m$\cdot$K).  The assembly is insulated by placing it inside an alumina tube.  As shown in Figure \ref{fig:heater}a, the glass cell is cemented (Omegabond 400) to a separate short (5 mm length) section of the nitride material designed to mate with the top of the spool.  The two-part design allows the fragile cell to be easily removed or reattached and kept separate during initial assembly or maintenance/repair of the heater.  The wire is wound as a twisted pair to minimize magnetic fields that could be produced when electrical current is applied.  

It is observed that cell heating in DC mode can present a major source of field noise in magnetometer measurements.  Although one can turn off the heater when measurements are made, this is unfavorable since the cell cools down during these times.  The preferred mode of heating is continuous AC, using a relatively high frequency ($>10$ kHz) far away from the signals of interest and outside the sensitive bandwidth (Figure \ref{fig:heater}b).  A type-T thermocouple is placed near the top of the aluminum nitride pillar, as close to the cell as possible, to measure the temperature and a proportional-integral-derivative temperature controller (Omega CN9000A model) provides a feedback loop to control the current applied to the heating wire.  The supply of alternating current is a low-distortion audio amplifier (AE Techron LVC2016, Crown Macro-Tech series or similar) where the AC input voltage is produced by a function generator and the amplitude is controlled by the feedback.  At a 40 kHz heating frequency a root-mean-square power of approximately 10 W is required to raise the cell temperature to 180 $^\circ$C.  

Thermal insulation of the vapor cell is an important design feature in the experimental setup.  While the cell is operated at 170-190 $^\circ$C to achieve the quoted sensitivities, it is desirable to detect NMR signals from the sample near to room temperature.  For insulation, an air or vacuum gap of 0.5-1 mm is left between the vapor cell and the bottom of the shuttling system.  The temperature of the air surrounding the NMR sample tube inside the shuttling system may also be regulated.  One must minimize temperature gradients across the cell to avoid further broadening of the atomic resonance line and a decrease in the overall sensitivity.  The minimization of thermal losses also allows the desired temperature to be reached at lower heating power.  

\begin{figure}%
\includegraphics[width=\columnwidth]{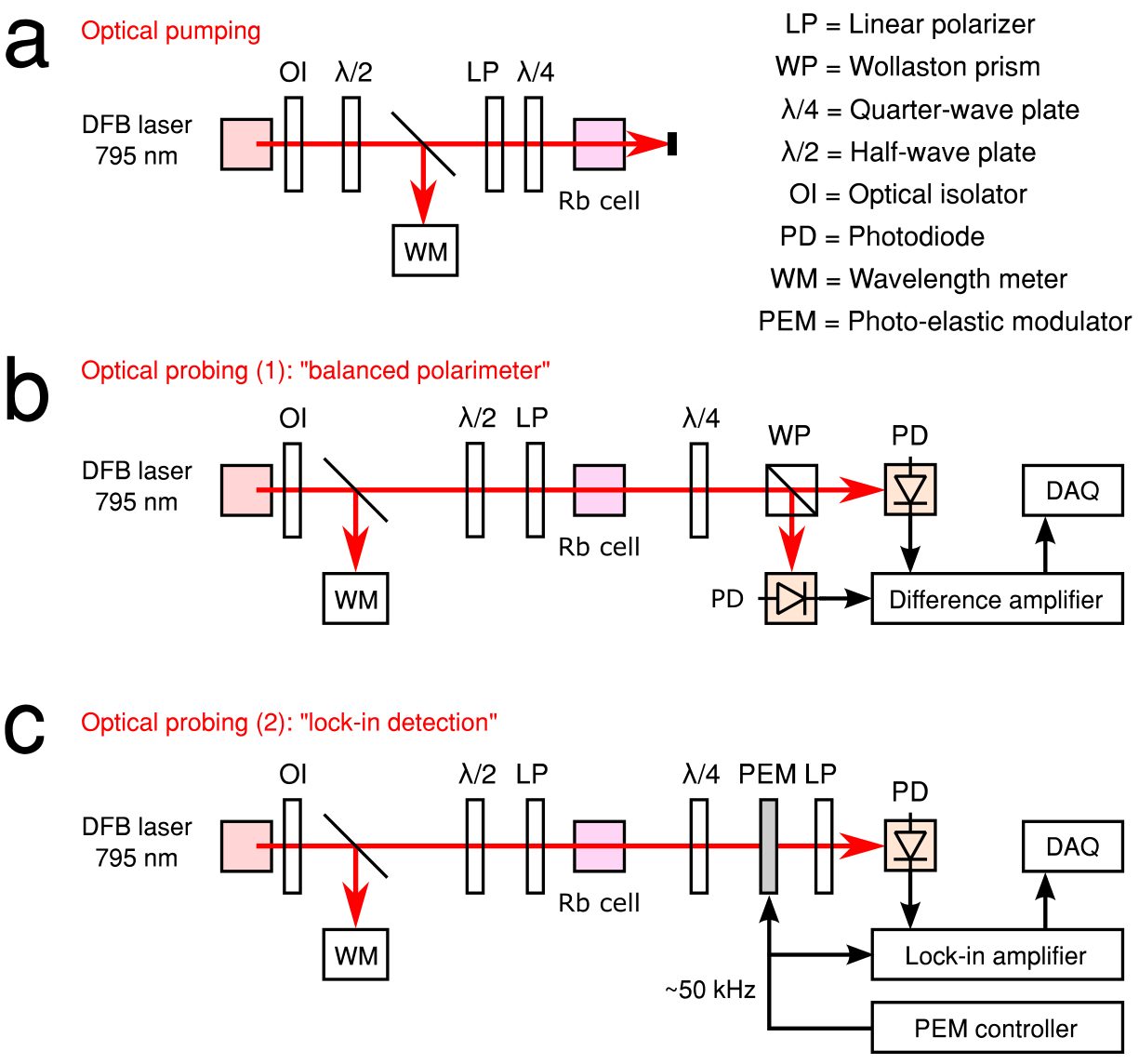}\vspace{0.5cm} \\
\includegraphics[width=\columnwidth]{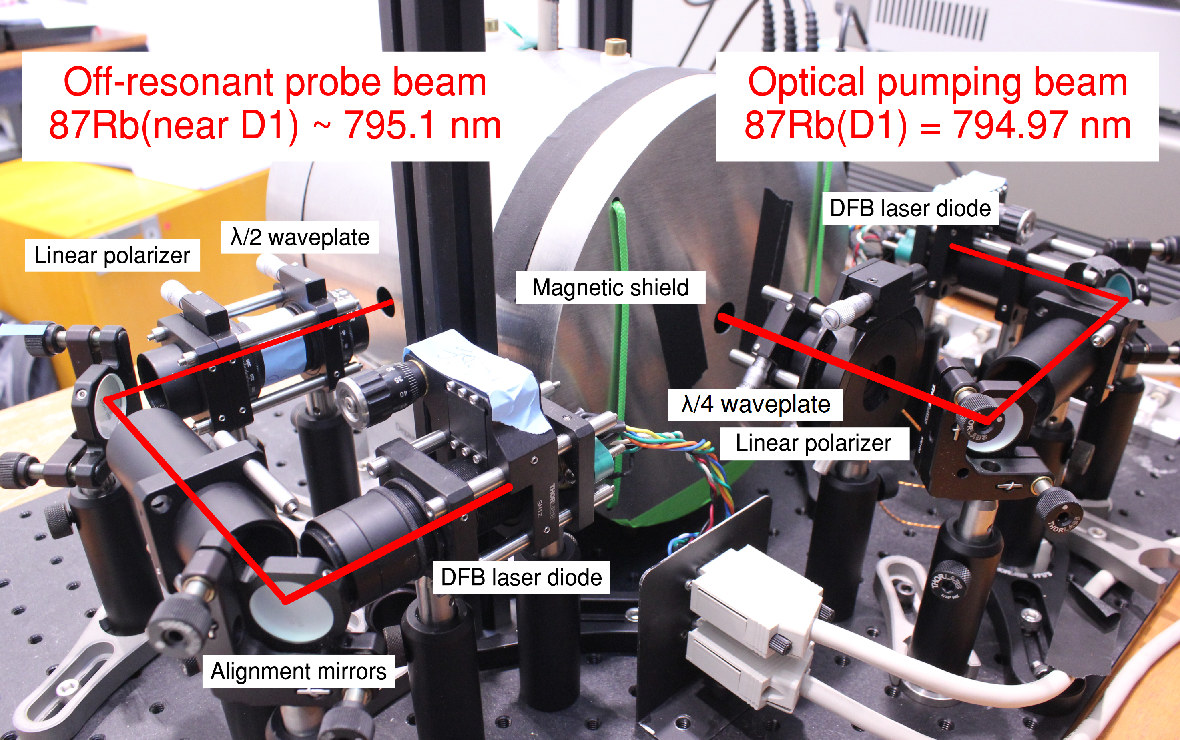}%
\caption{Layout of optical components used for signal detection: (a) light beam for optically pumping alkali metal spins; optical polarimetry using (b) quadrature or (c) heterodyne photo-detection.  The lower photograph illustrates the arrangement of (a) + (b) around the magnetic shield.}%
\label{fig:optics-schematic}%
\end{figure}

The optical setup is illustrated in Figure \ref{fig:optics-schematic}.  For the optical pumping of the $^{87}$Rb vapor a tunable distributed-feedback (DFB) diode laser is used with a Thorlabs ITC502 laser diode controller.  The laser light is tuned to the $^{87}$Rb D1 line at a wavelength of 794.970 nm\cite{SteckRb201} using a wavelength meter / interferometer (e.g.\ Agilent / HP 86120B, 0.005 nm accuracy).  An optical isolator is positioned immediately after the laser collimator to eliminate reflections from the subsequent optical components.  This initial beam is plane-polarized, so a half-wave plate is positioned in the path of propagation to allow arbitrary rotation of its plane of polarization.  A linear polarizer follows, to allow the light power to be attenuated to the desired level.  Transmitted light is then converted to circularly polarized light with a quarter-wave plate whose principal axes are 45$^\circ$ from the axis of the linear polarizer.  In normal operation, settings are chosen such that a power of approximately 15 mW arrives at the vapor cell.  In addition to the components that control the beam polarization, beam splitters or flip mirrors may be added to divert light into a variety of meters for tuning, locking the laser, or measuring the beam power.

The source of the probe beam is a second DFB laser diode controlled with a separate ITC502 unit and tuned near the D1 line.  For the most sensitive detection one should effect the largest optical rotation on the beam while minimizing resonant excitation of the atoms.  The apparatus achieves this condition when the probe laser is detuned between 1 and 2 times the linewidth (in our case, 100 to 150 GHz) away from the D1 frequency and when the power of the incident beam at the vapor cell is less than 10 mW.  Further optimization is described in Section \ref{sec:calibration}.  Two methods for measuring the optical rotation are outlined in Figures \ref{fig:optics-schematic}b and \ref{fig:optics-schematic}c.  Figure \ref{fig:optics-schematic}b illustrates a ``balanced polarimeter'', where after the alkali vapor imparts rotation, the probe beam is split by a Wollaston prism into perpendicular components of the polarization and the intensities are detected at a pair of photodiodes.  These are ``balanced'' (i.e.\ zeroed) by adjusting the incident beam's plane of polarization to equalize the detected intensities.  The photodiode voltages are subtracted and then amplified so the final output voltage is the response to a change in field and is proportional to the optical rotation angle.  It is useful to add a quarter-wave plate in the beam path to correct for additional optical rotation due to birefringent walls of the vapor cell.

The scheme in Figure \ref{fig:optics-schematic}c illustrates a method for heterodyne detection, which aims to minimize the influence of extremely-low-frequency noise coming from the apparatus such as $1/f$ noise and laser jitter.  The probe beam polarization is modulated at a frequency of around 50 kHz using a photoelastic modulator (PEM, Hinds Instruments model PEM-100).  The PEM consists of a birefringent quartz crystal that is stressed by mechanical vibration, induced with a piezoelectric element, along one of its principal axes to retard incident light by up to $\pm\lambda/4$.  If the incident light is linearly polarized at 45$^\circ$ to the fast axis of the crystal, the oscillating stress alternates the transmitted light between left and right circular polarizations at the natural resonance frequency of the PEM crystal.  This light passes through a linear polarizer, resulting in the intensity becoming modulated at the PEM frequency, which is then detected at a photodiode.  Demodulation against the PEM resonance frequency is performed with a lock-in amplifier (Stanford Research Systems model SR830) to give the final output signal.  The DC offset in this case can also be zeroed by rotating the quarter-wave plate.  The integration time of the lock-in (10 $\mu$s to 100 ms) is chosen as appropriate for the signal-frequency region of interest.

For both detection methods, careful alignment of the pump and probe beams is required both with respect to one another and the sample under study.  We find that the pump and the probe should be aligned perpendicular to each other within $\sim 2$ degrees.  Additionally, detection is most sensitive when the pump and probe beams (i) intersect over a large volume of the vapor cell, ideally its entire volume and (ii) are as close as possible to the NMR sample.

The output of the differential or lock-in amplifier may be connected to a spectrum analyzer for real-time continuous monitoring of the frequency spectrum and a voltmeter to measure the DC offset voltage.  To stabilize the DC offset a second-order high-pass RC circuit, with a cutoff frequency around 0.3 Hz, is placed immediately after the output of the amplifier.  

For NMR experiments, the magnetometer output is recorded digitally using a data acquisition card interfaced with a PC (National Instruments USB6229, 16-bit $\pm 5$ V analog input range).  A graphical-user-interface on the PC allows the instrument user to define experimental parameters including spectral bandwidth, acquisition time and the number of dwell points, and display the Fourier transform of the recorded data.  The program also interfaces the analog and digital I/O needed to operate other parts of the spectrometer, such as setting the waveforms that control the pulse sequences and sample shuttling.

\subsection{Calibrations}
\label{sec:calibration}
When setting up the magnetometer a number of optimization and calibration procedures should be performed.  The field in the vicinity of the vapor cell and the NMR sample must be controlled, in order to (1) operate in the sensitive SERF regime and (2) set known values of the bias fields.  The magnetometer signal as a function of frequency must then be measured to determine detection bandwidth.  The absolute sensitivity of the magnetometer should also be determined and checked on a regular basis to assess the condition of the vapor cell.

The magnetometer sensitivity is measured by applying an alternating field on the order of 1 pT to 1 nT through one of the $x$, $y$, or $z$ field coils.  The field is supplied using a function generator connected to the coils in series with a 1 M$\Omega$ to 1 k$\Omega$ shunt resistor, which produces alternating current in the 1 $\mu$A to 1 mA range.  The oscillating magnetic field along the corresponding axis around the vapor cell generates a magnetometer response that we call the ``test signal''.  

\begin{figure}%
\includegraphics[width=\columnwidth]{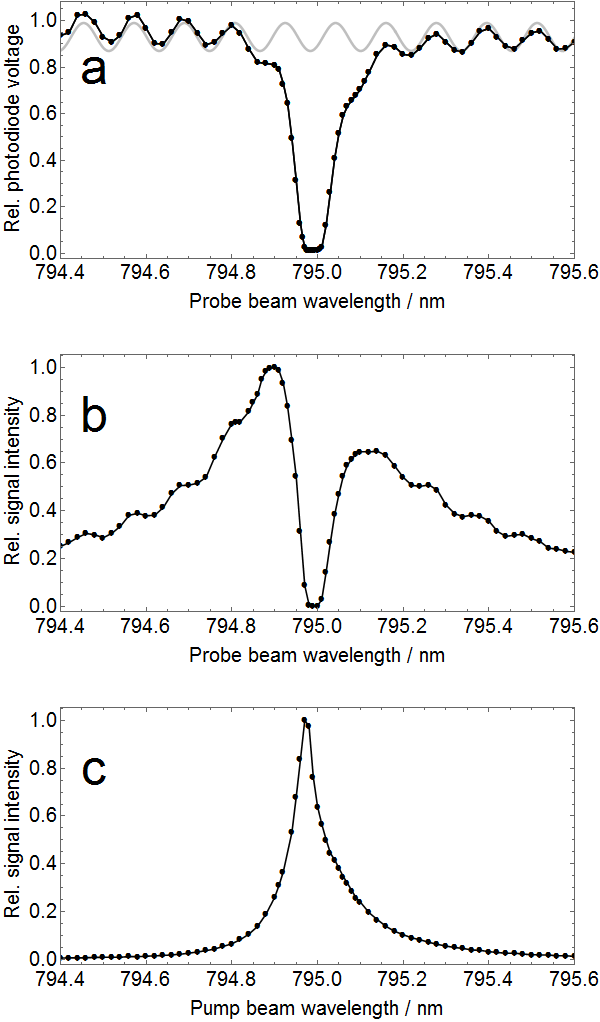}
\caption{(a) Optical transmission of 10 mW laser light through the $^{87}$Rb vapor cell at 175 $^\circ$C showing a broad absorbance peak (0.1 nm width at half max) at the $^{87}$Rb D1 transition.  plots (b) and (c) show, respectively, the magnetometer response to a 20 pT, 100 Hz test signal as a function of pump and probe beam wavelengths.  The incident probe beam power is 10 mW and the power of the optical pumping beam is 20 mW; in (b) the pump beam wavelength is held constant at 794.97 nm while in (c) the probe beam wavelength is at 794.88 nm.  Periodic fringe peaks in (a) and (b), highlighted by the gray curve, are consistent with interference due to a Fabry-Perot effect at the glass walls of the cell (total thickness $\approx$ 1.6 mm).}%
\label{fig:testsignal-pumpprobefrequency}%
\end{figure}

To observe the test signal, the pump laser beam is tuned near to the D1 transition frequency.  Using a power meter, the transmitted power of the probe beam is measured versus frequency.  For our vapor cell, we expect and observe a Voigt absorption profile with about 50 GHz (0.1 nm) width as shown in Figure \ref{fig:testsignal-pumpprobefrequency}a. Following this, the probe laser beam is tuned approximately twice the linewidth from the D1 line on the high-frequency side.  At the same time the pump laser is tuned to the center of the optical transition and the power level set to 10 mW.  These ``crude'' settings should allow a test signal amplitude of 1 nT to be detected even at relatively low temperatures of the alkali gas, around 120 $^\circ$C for $^{87}$Rb.  To fully optimize the performance of the magnetometer one should refine the alignment, power and frequency of both pump ($\pumpaxis$ axis) and probe ($\probeaxis$ axis) laser beams until the signal-to-noise ratio of the test signal along the $\sensaxis$ axis is maximized.  Sample data for our magnetometer are shown in Figure \ref{fig:testsignal-pumpprobefrequency}(b) and (c).

The test signal is also used to find the values of the bias currents that cancel remaining magnetic fields.  From Equation \ref{eq:shimxyz} if a slowly oscillating ($< 100$ Hz) test signal is applied in the perpendicular direction to both pump and probe, the magnetometer response is a dispersive Lorentzian function.  The residual $B_\sensaxis$ is zeroed first.  The pump laser beam is blocked from reaching the cell and the polarization of the probe beam is rotated until the magnetometer response becomes zero.  The pump beam is then unblocked and $B_z$ adjusted to zero the DC offset of the magnetometer.  To eliminate residual fields $B_\probeaxis$ or $B_\pumpaxis$, a $\approx 5$ Hz test signal is applied in turn along the $\pumpaxis$ or $\probeaxis$ axes, respectively, viewing the magnetometer response on an oscilloscope.  A test signal oscillating parallel to the $\probeaxis$ axis is applied and $B_\pumpaxis$ is adjusted until the response is minimized.  The field $B_\probeaxis$ is then adjusted to minimize the response from a test field oscillating along the $\pumpaxis$ axis.  Fields $B_x$, $B_y$ and $B_z$ are adjusted iteratively in this way until the test signal is only detected along $z$.  Unfortunately, this protocol does not guarantee a zero field at either the magnetometer or the NMR sample.  Finite magnetic fields can arise unintentionally for several reasons: the NMR sample is necessarily located at some distance from the vapor cell, meaning gradients in the shimming fields can cause a nonzero field at the sample.  Also, the test signal does not distinguish $B_x$, $B_y$ or $B_z$ from fictitious magnetic fields generated by light shifts, or imperfect alignment of the laser beams.  For further accuracy, the fields should be minimized by measuring NMR spectra to determine the values of field from the Larmor frequency.  The field per unit current and the zero bias fields can then be calculated.

We quantify the sensitivity by the smallest test signal amplitude $\delta B_\sensaxis$ that is detectable with unit signal-to-noise ratio (SNR) over a 1 Hz measurement bandwidth.  The SNR is defined at the peak amplitude divided by the standard deviation of the noise.  On our setup, a 100 Hz, 4.4 pT test signal yields an SNR of around 100 when sampled over 0.5 s ($=1$ Hz bandwidth).  The resulting sensitivity is therefore $\delta B_\sensaxis \approx 30$ fT/Hz$^{1/2}$ at 100 Hz. 

\begin{figure}%
\includegraphics[width=0.8\columnwidth]{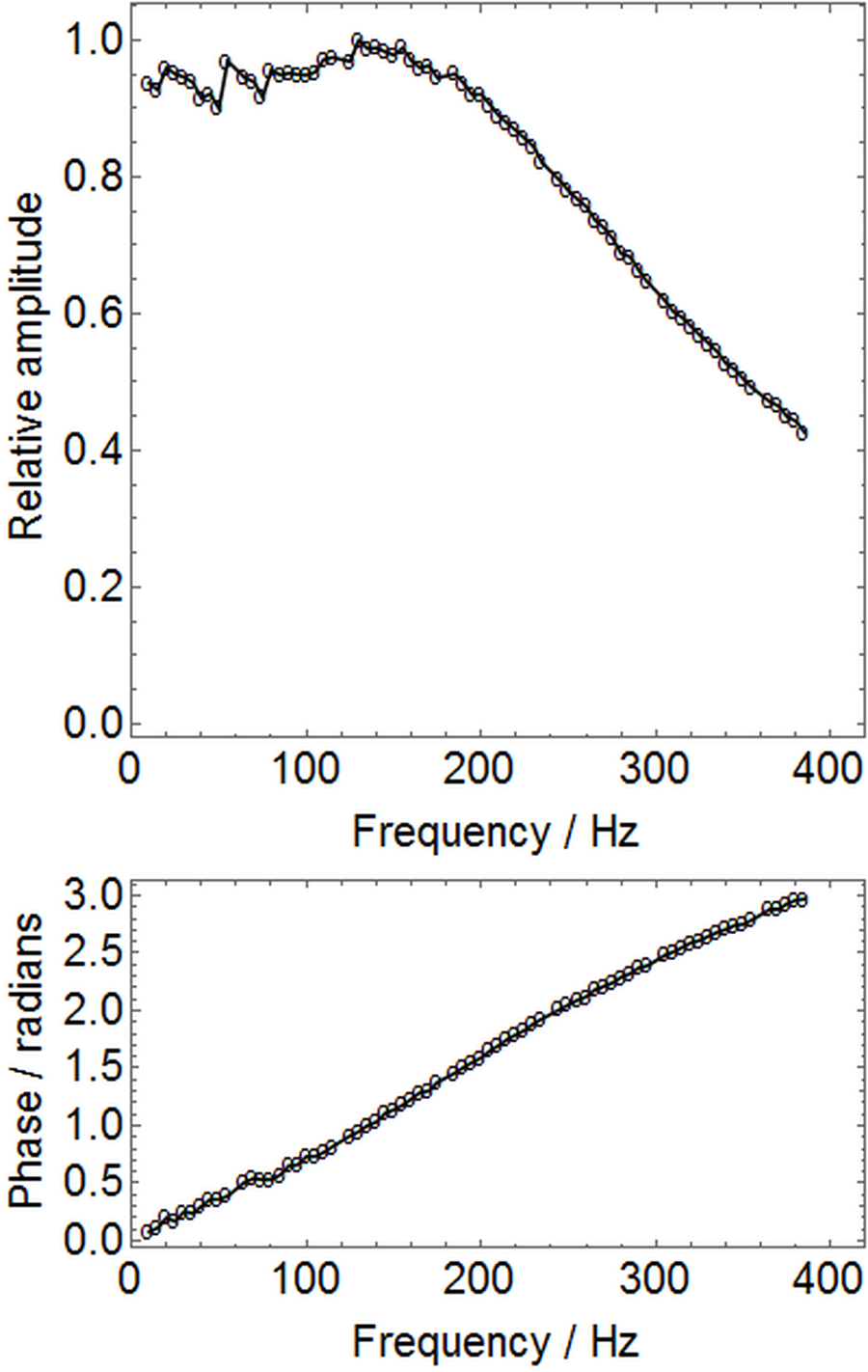}%
\caption{Plots showing amplitude and phase of the magnetometer signal in response to a 4 pT sinusoidal test signal, as a function of the test signal frequency.}%
\label{fig:testsignalvsfrequency}%
\end{figure}

The bandwidth of the magnetometer is determined from a set of spectra measured at different frequencies of the test signal.  Figure \ref{fig:testsignalvsfrequency} shows magnetometer signal amplitude and phase relative to a 4.4 pT test signal over the frequency range of $0-400$ Hz.  The data show that the signal amplitude drops by close to 65 percent and the phase shift (relative to zero frequency) varies from zero at 0 Hz to almost 1.3 radians at 400 Hz.  These data suggest a usable magnetometer bandwidth of approximately 300 Hz.  However, if the noise is dominated by sources from outside the cell, for example Johnson noise from the shields and not on characteristics of the Rb vapor, then the noise is attenuated in the same way and the signal-to-noise ratio does not change strongly across the spectrum, so the usable frequency range is approximately a factor of 2-3 larger.  We find that SNR of the magnetometer is essentially flat up to 500 Hz, with the exception of vibrational noise that can degrade sensitivity in the spectral region below 10 Hz.  These parameters vary depending on alkali vapor composition, temperature, pump beam power and other parameters that influence the resonance linewidth of the alkali-atom spins.

Although the magnetometer settings are chosen as a compromise between sensitivity and detection bandwidth (inverse of response time), there are some aspects of the NMR experiment that would benefit from a change of conditions.  One of these is the time to recover a magnetometer signal following pulsed fields of several mT submitted by the NMR pulse sequences, which on our setup can be as large as 20-30 ms.  In a straightforward pulse-acquire experiment this dead time of the magnetometer does not cause problems; it is sufficient to omit the data points acquired 30-40 ms after pulsed fields and correct for the time shift via a first-order (frequency-linear) phase correction to the spectrum.  However, the response time is generally too long to allow advanced experimental techniques, such as spin decoupling, that involve so-called ``pulse-windowed detection''.  The aim of such methods is to acquire signal data points during short delays in between magnetic field pulses, which may only be a few ms.  To reduce the magnetometer dead time without compromising the detection sensitivity, one should be able to rapidly switch the intensity of the pump beam.  Following application of pulsed magnetic fields, the intensity should be increased (thereby temporarily increasing $R_\mathrm{OP}$ and decreasing $T_2$) to accelerate magnetometer recovery before being returned to the level which allows the NMR signal to be recorded.

Most of our ZULF-NMR experiments are performed with liquid samples whose volume is on the order of 0.05 to 0.5 mL.  The liquids are contained in standard 5 mm or 10 mm o.d.\ glass NMR tubes.  The NMR signal is detected with the tube positioned just above the alkali vapor cell.  Since the magnetic field generated by the sample decreases with the third power of distance, flat-bottomed thin-wall NMR tubes are preferred, since they reduce the distance between the sample and vapor cell and lead to improved sensitivity.  For the 5 mm o.d.\ tube, the sensitive volume is the lowest 5-10 mm of the liquid corresponding to volumes 100-200 $\mu$L.  The magnetometer is also suitable for use with microfluidic devices or flow cells,\cite{Savukov2005PRL,Ledbetter2008PNAS105,Ledbetter2009JMR199} where the volume of interest is placed as close as possible to the vapor cell.

A preferred sample for calibration and sensitivity determination is neat $99\%$ $[1-^{13}$C$]$-formic acid (H$^{13}$COOH, 26.5 mol/dm$^3$; zero-field NMR spectrum peak at $^{1}J_{\mathrm{CH}} \approx 220$ Hz).   Another choice of sample is $99\%$ [$^{13}$C]-methanol ($^{13}$CH$_3$OH, 24.7 mol/dm$^3$; zero-field NMR spectral peaks at $^{1}J_{\mathrm{CH}} \approx 140$ Hz and $2\times$ $^{1}J_{\mathrm{CH}}$).  Under the quoted conditions the zero-field NMR signals of these samples are in excess of 100 times the noise floor, corresponding to fields on the order of 1 pT.  For best results the liquids should be degassed (freeze-pump-thaw) prior to use, in an effort to maximize the natural lifetime of the spin coherences by eliminating dissolved paramagnetic oxygen (O$_2$).

Finally, the sensitivity of the magnetometer is also strongly dependent on bias field.  This is expected since the relaxation time of the alkali spins depends on field, particularly when bias fields are strong enough to withdraw SERF behavior (around 100 nT).  Additionally there is a dependence because the large difference in gyromagnetic ratio of the alkali-atom spins and the NMR sample leads to a reduced overlap of the $^{87}$Rb resonance with the NMR transition frequencies.  This behavior can be seen upon solving Equation \ref{eqn:blochalkali} for an oscillating field along the $\sensaxis$ axis, $\bm{B} = \bm{e_\sensaxis} B_1 \cos(2\pi\nu_1 t)$ in the presence of a bias field $\bm{e_\pumpaxis} B_\pumpaxis$ along a perpendicular axis.  The maximum response is shifted to a non-zero field defined by $\nu_\pumpaxis = \mu_B g B_\pumpaxis / h q$:\cite{BudkerOpticalMagnetometry}
\begin{eqnarray}
S_\probeaxis &=& (S_0/2) B_1 \Delta\nu \times \nonumber \\ 
&&  \Bigl[ \frac{\Delta\nu \cos(2\pi\nu_1 t) + (\nu_1 - \nu_\pumpaxis)\sin(2\pi\nu_1 t)}{\Delta\nu^2 + (\nu_1 - \nu_\pumpaxis)^2} 
+ \nonumber \\ 
&& \qquad \frac{\Delta\nu \cos(2\pi\nu_1 t) - (\nu_1 - \nu_\pumpaxis)\sin(2\pi\nu_1 t)}{\Delta\nu^2 + (\nu_1 - \nu_\pumpaxis)^2} \Bigr], \label{eq:biasfieldsensitivity}
\end{eqnarray}
where $\Delta\nu = 1/T_2$.  

\begin{figure}%
\includegraphics[width=\columnwidth]{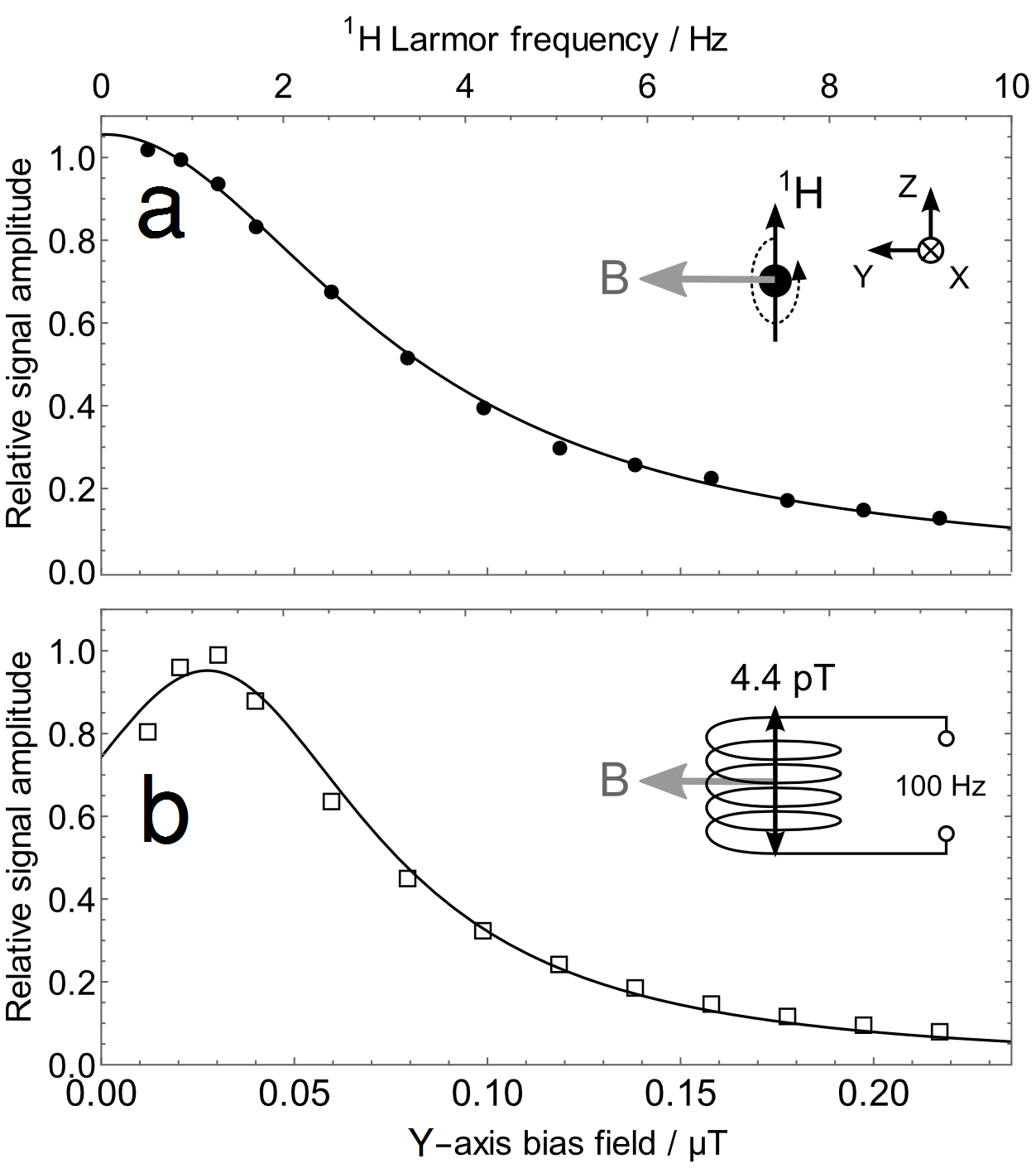}%
\caption{Magnetometer sensitivity plotted measured as a function of a magnetic bias field along the $\pumpaxis$ axis.  Data points represent the amplitude of the detected signals originating from: (a, circles $\bullet$) $^1$H spin precession in a 0.1 mL sample of tap water; (b, squares $\square$) a 4.4 pT, 100 Hz test signal oscillating along the $\sensaxis$ axis.  The magnetometer detects the component of field along the $\sensaxis$ axis.  Curve shapes are consistent with Equation \ref{eq:biasfieldsensitivity} and explained in the main text.}%
\label{fig:biasfields}
\end{figure}

The data in Figure \ref{fig:biasfields} illustrate the sensitivity of our magnetometer to small oscillating fields on the order of pT in the presence of bias fields up to around 0.25 $\mu$T.  Figure \ref{fig:biasfields}a shows the intensity of the NMR signal for $^1$H Larmor precession in water over the frequency range 1-10 Hz (approximately 0 to 0.25 $\mu$T).  For each data point it was necessary to (approximately) compensate for the much stronger magnetometer response to the bias field, so that the voltage of the photodiodes did not ``clip'' out of the range where they could be measured.  The overall profile fits to an absorptive Lorentzian lineshape centered near zero frequency, consistent with Equation \ref{eq:biasfieldsensitivity} for $|\nu_1| = |\gamma_\mathrm{H}B_\pumpaxis/2\pi| \ll |\nu_\pumpaxis|$.  Although there is a rapid loss of sensitivity with increasing bias field, the Larmor precession is easily detected over the 0-10 Hz region, enabling us to execute the field-zeroing protocol described earlier in this section and the near-zero-field NMR spectroscopy described in Section \ref{sec:NMRspindynamics}.  The data in Figure \ref{fig:biasfields}b represent the magnetometer response to a 4.4 pT, 100 Hz test field over the same range of bias fields.  The center of the Lorentzian is shifted away from zero field, since the magnitude of the signal frequency $\nu_1$ is now comparable with that of $\nu_y$.   

\section{Summary and outlook}
We have described the setup and characterization of an instrument for detecting NMR signals in magnetic fields below $\approx10^{-7}$ T, which we refer to as the ZULF regime.  The key stages of the NMR experiment were reviewed: sample prepolarization, signal encoding using pulsed magnetic fields and finally optical detection via resonant non-linear magneto-optical rotation near the $^{87}$Rb D1 transition in a SERF $^{87}$Rb-vapor magnetometer.  The instrument is used for direct measurements of nuclear spin-spin couplings and Larmor precession near zero field.  Indirect measurements of NMR up to fields of a few mT can also be made with high sensitivity, which includes NMR in the earth's field.\cite{Ganssle2014ANIE53}  The whole instrument is relatively inexpensive, compact and easy to maintain compared to traditional NMR apparatus.  As the techniques based on SERF magnetometers become mature, we expect the practical realization of many promising NMR applications, where the sensitivity or cost of low-field detection has hitherto been prohibitive.  Examples include the characterization of liquids confined in pores, porous metals,\cite{Xu2008JMRI28} emulsions and other inhomogeneous materials, the proposed measurement of antisymmetric spin-spin-coupling tensors in chiral molecules\cite{King2016XX} and the use of spin decoupling techniques for multidimensional spectroscopy.\cite{Sjolander2017JPCLXX}

Although SERF alkali magnetometry techniques continue to be refined, fundamental sensitivity limits are close to being reached.  At this stage, the mass sensitivity of ZULF NMR still remains several orders of magnitude below the state-of-the-art portable NMR instrumentation based on coil induction even at signal frequencies as low as 1 MHz.  Room-temperature pre-polarization of samples around 1-2 T does not produce sufficient signal intensity for ZULF NMR to be used for routine analyses of organic molecules in mM concentration, even with full $^{13}$C isotopic enrichment.  However, this limitation can be overcome by techniques that produce a much higher starting polarization, for instance parahydrogen-induced polarization.\cite{Theis2011NatPhys7,Theis2012JACS134,Butler2013JCP138}  Dynamic nuclear polarization (DNP) is another experimental strategy for producing order-of-unity nuclear spin polarization in liquids, promising signal strength enhancement by a factor 10$^5$-10$^6$.\cite{Abragam1978RPP41,WenckebachBOOK2016,Kockenberger2014emagres}  Another interesting future direction of ZULF NMR is extending the techniques to single-molecule level using magnetometers based on single NV centers in diamond that have already demonstrated single-spin NMR capabilities, though not at ZULF.\cite{Lovchinsky2016Science351} This will take the technique to an unbeatable range of mass sensitivity and is ideal for stochastic polarization.  The same may be true for magnetic resonance imaging (MRI).  We note that only one decade since the first demonstration of ZULF NMR with atomic magnetometers,\cite{Savukov2005PRL} several works have demonstrated MRI, assisted by remote detection,\cite{Xu2008JMRI28,Xu2006PNAS103,Xu2008PRA78} extended-cell and flux-transformer strategies,\cite{Savukov2013JMR231,Savukov2013JMR233} although the alkali-vapor magnetometer and/or spatial encoding is used at higher magnetic fields on the order of mT and not in the ZULF/SERF regime.  At present, SQUID magnetometry is preferred for larger spatial fields-of-view, (e.g.\ imaging medical/physiological subjects), mainly because the technology is more mature.\cite{Clarke2007ARBE9,Mossle2006JMR179,KrausJrbook}  As the alkali-vapor, SQUID and NV-diamond magnetometers continue to undergo a rapid stage of development and advance, including in ZULF, it is expected that the scenes will change dramatically over the coming years.


\section*{Acknowledgement}
We are grateful to M. P. Ledbetter for his critically important contribution in the development of the apparatus and techniques described in this paper.  This material is based upon work supported by the National Science Foundation under Grant No.\ CHE-1308381 (authors DB and AP) and by the European Commission under the Marie Curie International Outgoing Fellowship Programme under Grant Agreement FP7-625054 ODMR-CHEM (author MCDT).  The opinions, findings, and conclusions or recommendations expressed in this material are those of the authors and do not necessarily reflect the views of the National Science Foundation, Cambridge University or the European Commission.


\begin{thebibliography}{0}%
\makeatletter
\providecommand \@ifxundefined [1]{%
 \@ifx{#1\undefined}
}%
\providecommand \@ifnum [1]{%
 \ifnum #1\expandafter \@firstoftwo
 \else \expandafter \@secondoftwo
 \fi
}%
\providecommand \@ifx [1]{%
 \ifx #1\expandafter \@firstoftwo
 \else \expandafter \@secondoftwo
 \fi
}%
\providecommand \natexlab [1]{#1}%
\providecommand \enquote  [1]{``#1''}%
\providecommand \bibnamefont  [1]{#1}%
\providecommand \bibfnamefont [1]{#1}%
\providecommand \citenamefont [1]{#1}%
\providecommand \href@noop [0]{\@secondoftwo}%
\providecommand \href [0]{\begingroup \@sanitize@url \@href}%
\providecommand \@href[1]{\@@startlink{#1}\@@href}%
\providecommand \@@href[1]{\endgroup#1\@@endlink}%
\providecommand \@sanitize@url [0]{\catcode `\\12\catcode `\$12\catcode
  `\&12\catcode `\#12\catcode `\^12\catcode `\_12\catcode `\%12\relax}%
\providecommand \@@startlink[1]{}%
\providecommand \@@endlink[0]{}%
\providecommand \url  [0]{\begingroup\@sanitize@url \@url }%
\providecommand \@url [1]{\endgroup\@href {#1}{\urlprefix }}%
\providecommand \urlprefix  [0]{URL }%
\providecommand \Eprint [0]{\href }%
\providecommand \doibase [0]{http://dx.doi.org/}%
\providecommand \selectlanguage [0]{\@gobble}%
\providecommand \bibinfo  [0]{\@secondoftwo}%
\providecommand \bibfield  [0]{\@secondoftwo}%
\providecommand \translation [1]{[#1]}%
\providecommand \BibitemOpen [0]{}%
\providecommand \bibitemStop [0]{}%
\providecommand \bibitemNoStop [0]{.\EOS\space}%
\providecommand \EOS [0]{\spacefactor3000\relax}%
\providecommand \BibitemShut  [1]{\csname bibitem#1\endcsname}%
\let\auto@bib@innerbib\@empty
\end{thebibliography}%


\section*{References}

\begin{thebibliography}{99} \label{sec:TeXbooks}


\bibitem{Hill2011emagres} Hill, H. D. W. and Gray, G. A., Spectrometers: a general overview, \emph{eMagRes} (2011).

\bibitem{Sykora2005ebyte} S\'{y}kora, S., NMR sensitivity: novel approaches and perspectives, DOI: 10.3247/SL1Nmr05.003 (web article, 2005).

\bibitem{Iriguchi1993JAP73} Iriguchi, N., The power sensitivity of magnetic resonance experiments, \emph{J. Appl. Phys.} 73, 2956-2957 (1993).

\bibitem{Abragambook1961} Abragam, A., \emph{The principles of nuclear magnetism}. Oxford, Clarendon Press. (1961).

\bibitem{Hoult1976JMR24} Hoult, D. I. and Richards, R. E., The signal-to-noise ratio of the nuclear magnetic resonance experiment, \emph{J. Magn. Reson.} 24, 71-85 (1976).

\bibitem{Hoult1979JMR34} Hoult, D. I. and Lauterbur, P. C., The sensitivity of the zeugmatographic experiment involving human samples, \emph{J. Magn. Reson.} 34, 425-433 (1979).

\bibitem{Hoult2007EMR} Hoult, D. I., \emph{Sensitivity of the NMR Experiment}, eMagRes, John Wiley and Sons, Ltd (2007).

\bibitem{Freeman1991CR91} Freeman, R., Selective excitation in high-resolution NMR, \emph{Chem. Rev.} 91, 1397-1412 (1991).

\bibitem{Styles1984JMR60} Styles, P. and Soffe, N. F., A high-resolution NMR probe in which the coil and preamplifier are cooled with liquid helium, \emph{J. Magn. Reson.} 60, 397-404 (1984).

\bibitem{Martin2005ARNMR56} Martin, G. E., Small-volume and high-sensitivity NMR probes, \emph{Ann. Rep. NMR Spectrosc.} 56, 1-96 (2005).

\bibitem{Lacey1999CR99} Lacey, M. Subramanian, R., Olson, D. L., Webb, A. G. and Sweedler, J. V., High-resolution NMR spectroscopy of sample volumes from 1 nL to 1 $\mu$L, \emph{Chem. Rev.} 99, 3133-3152 (1999).



\bibitem{McDermott2002Science295} McDermott, R., Trabesinger, A. H., Muck, M., Hahn, E., Pines, A. and Clarke, J., Liquid-state NMR and scalar couplings in micro-tesla magnetic fields, \emph{Science} 295, 2247-2249 (2002).

\bibitem{Thayer1987ACR20} Thayer, A. M. and Pines, A., Zero-field NMR, \emph{Acc. Chem. Res.} 20, 47--53 (1987).

\bibitem{Zax1985JCP83} Zax, D. B., Bielecki, A., Zilm, K., W., Pines, A. and Weitekamp, D., Zero-field NMR and NQR, \emph{J. Chem. Phys.} 83, 4877-4905 (1985).

\bibitem{BlanchardPRL2015} Blanchard, J. W., Sjolander, T. F., King, J. P., Ledbetter M. P., Levine E. H., Bajaj V. S., Budker D. and  Pines A., Measurement of untruncated nuclear spin interactions via zero- to ultralow-field nuclear magnetic resonance, \emph{Phys. Rev. B} 92, 220202(R) (2015).

\bibitem{Mossle2006JMR179} M{\"o}{\ss}le, M., Han, S.-I., Myers, W. R., Lee, S.-K., Kelso, N., Hatridge, M., Pines, A. and Clarke, J., {SQUID}-detected microtesla {MRI} in the presence of metal, \emph{J. Magn. Reson.} 179, 146-151 (2006).

\bibitem{Freedman2014RSI85} Freedman, R., Anand, V., Grant, B., Ganesan, K., Tabrizi, P., Torres, R., Catina, D., Ryan, D., Borman, C. and Crueckl, C., A compact high-performance low-field {NMR} apparatus for measurements on fluids at very high pressures and temperatures, \emph{Rev. Sci. Instrum.} 85, 025102 (2014).

\bibitem{Emondts2014PRL112} Emondts, M., Ledbetter, M. P., Pustelny, S., Theis, T., Patton, B., Blanchard, J. W., Butler, M., Budker, D. and Pines, A., Long-lived heteronuclear spin singlet states in liquids at a zero magnetic field, \emph{Phys. Rev. Lett.} 112, 077601, (2014).

\bibitem{Blanchard2016emagres} Blanchard, J. W. and Budker, D., Zero- to ultralow-field {NMR}, \emph{eMagRes} accepted manuscript (2016).

\bibitem{Ledbetter2009JMR199} Ledbetter, M. P., Crawford, C. W., Pines, A., Wemmer, D. E., Knappe, S., Kitching, J. and Budker, D., Optical detection of NMR J-spectra at zero magnetic field, \emph{J. Magn. Reson.} 199, 25-29 (2009).

\bibitem{Ledbetter2011PRL107} Ledbetter, M. P., Theis, T., Blanchard, J. W., Ring, H., Ganssle, P. J., Appelt, S., Blumich, B., Pines, A. and Budker, D., Near-zero-field nuclear magnetic resonance, \emph{Phys. Rev. Lett.}, 107, 107601 (2011).

\bibitem{Ledbetter2013PT66} Ledbetter, M. P. and Budker, D., Zero-field nuclear magnetic resonance, \emph{Physics Today}, 66, 44 (2013).

\bibitem{Butler2013JCP1382} Butler, M., Ledbetter, M. P., Theis, T., Blanchard, J. W., Budker, D. and Pines, A., Multiplets at zero magnetic field: the geometry of zero-field NMR, \emph{J. Chem. Phys.} 138, 184202-15 (2013).

\bibitem{Blanchard2013JACS} Blanchard, J. W., Ledbetter, M. P., Theis, T., Butler, M., Budker, D. and Pines, A., High-resolution zero-field NMR J-spectroscopy of aromatic compounds, \emph{J. Am. Chem. Soc.} 135, 3607-3612 (2013).

\bibitem{Theis2013CPL580} Theis, T., Blanchard, J. W., Butler, M., Ledbetter, M. P., Budker, D. and Pines, A., Chemical analysis using J-coupling multiplets in zero-field NMR, \emph{Chem. Phys. Lett.}, 580, 160-165 (2013).





\bibitem{Lenz1990IEEEProc78} Lenz, J. E., A review of magnetic sensors, \emph{Proc. IEEE} 78, 973-989 (1990).

\bibitem{Kominis2003Nature422} Kominis, I. K., Kornack, T. W., Alfred, J. C. and Romalis, M. V., A subfemtotesla multichannel atomic magnetometer, \emph{Nature} 422, 596-599, (2003). 

\bibitem{Savukov2005PRL} Savukov, I. M. and Romalis, M. V., NMR detection with an atomic magnetometer, \emph{Phys. Rev. Lett.} 94, 123001 (2005).

\bibitem{Savukov2007JMR185} Savukov, I. M., Seltzer, S. J. and Romalis, M. V., Detection of {NMR} signals with a radio-frequency atomic magnetometer, \emph{J. Magn. Reson.} 185, 214-220 (2007).

\bibitem{Ledbetter2008PRA77} Ledbetter, M. P., Savukov, I. M., Acosta, V. M., Budker, D. and Romalis, M. V., Spin-exchange-relaxation-free magnetometry with Cs vapor, \emph{Phys. Rev. A}, 77, 033408 (2008). 

\bibitem{Bevilacqua2009JMR201} Bevilacqua, G., Biancalana, V., Dancheva, Y. and Moi, L., All-optical magnetometry for NMR detection in a micro-tesla field an unshielded environment, \emph{J. Magn. Reson.} 201, 222-229 (2009).



\bibitem{Tiporlini2013TSW} Tiporlini, V. and Alameh, K., {High sensitivity optically pumped quantum magnetometer}, \emph{The Scientific World}, 858379 (2013).

\bibitem{Budker2007NatPhys3} Budker, D. and Romalis, M. V., Optical magnetometry, \emph{Nat. Phys.} 3, 227-234 (2007).

\bibitem{Budker2002RMP74} Budker, D., Gawlik, W., Kimball, D. F., Rochester, S. M., Yashchuk, V. V. and Weis, A., Resonant nonlinear magneto-optical effects in atoms, \emph{Rev. Mod. Phys.} 74, 1153–1201 (2002).

\bibitem{BudkerOpticalMagnetometry} \emph{Optical Magnetometry}, Eds. Budker, D. and Jackson Kimball, D. F., Cambridge University Press (2013). ISBN 1107010357.

\bibitem{Seltzerthesis} Seltzer, S. \emph{Developments in alkali-metal atomic magnetometry}, Ph.D. thesis, Princeton University (2008).

\bibitem{Happer1973PRL31} Happer, W. and Tang, H.; Spin-exchange shift and narrowing of magnetic resonance lines in optically pumped alkali vapors \emph{Phys. Rev. Lett.} 31, 273-276 (1973).

\bibitem{Happer1977PRA16} Happer, W. and Tam, A. C., Effect of rapid spin exchange on the magnetic resonance spectrum of alkali vapors, \emph{Phys. Rev. A}, 16, 1877-1891 (1977).

\bibitem{Savukov2005PRA71} Savukov, I. M. and Romalis, M. V., Effects of spin-exchange collisions in a high-density alkali-metal vapor in low magnetic fields, \emph{Phys. Rev. A} 71, 023405 (2005).

\bibitem{Allred2002PRL89} Allred, J. C., Lyman, R. N., Kornack, T. W. and Romalis, M. V., High-sensitivity atomic magnetometer unaffected by spin-exchange relaxation, \emph{Phys. Rev. Lett.} 89, 130801 (2002).

\bibitem{Greenberg1998RMP70} Greenberg, Y. S., Application of superconducting quantum interference devices to nuclear magnetic resonance, \emph{Rev. Mod. Phys.} 70, 175-222 (1998).

\bibitem{Friedman1986RSI} Friedman, L. J., Wennberg, A. K. M., Ytterboe, S. N. and Bozler, H. M., Direct detection of low-frequency {NMR} using a dc {SQUID}, \emph{Rev. Sci. Instrum.} 57, 410 (1986).

\bibitem{Qiu2009IEEETAS22} Qiu, L. Q., Zhang, Y., Krause, H. J., Braginski, A. I., Tanaka, S. and Offenhausser, A., High-Performance Low-Field NMR Utilizing a High- $T_{\rm c}$ rf SQUID, \emph{IEEE Transactions on Applied Superconductivity}, 19, 831-834, (2009).

\bibitem{Augustine1998SSNMR11} Augustine, M. P., TonThat, D. M. and Clarke, J., {SQUID} detected {NMR} and {NQR}, \emph{Solid State NMR} 11, 139-156 (1998).

\bibitem{Fan1989IEEETM2} Fan, N. Q., Heaney, M. B., Clarke, J., Newitt, D., Wald, L. L., Hahn, E. L., Bielecki, A. and Pines, A., Nuclear magnetic resonance with dc {SQUID} preamplifiers, \emph{IEEE Trans. Magn.} 25, 1193-1199 (1989).

\bibitem{KrausJrbook} Kraus Jr., R. H., Espy, M., Magnelind, P. and Vogelov, P., \emph{Ultra-low-field nuclear magnetic resonance: a new {MRI} regime}, Oxford University Press, New York (2014).

\bibitem{Espy2005IEEETAS15} Espy, M., Matlachov, A. N., Vogelov, P. and Kraus Jr., R. H., {SQUID}-based simultaneous detection of {NMR} and biomagnetic signals at ultra-low magnetic fields, \emph{IEEE Trans. Appl. Supercond.} 15, 635--639 (2005).

\bibitem{Matlashov2011IEEEApplSupercon21} Matlashov, A. N., Schultz, L. J., Espy, M. A., Kraus, R. H., Savukov, I. M., Volegov, P. L., and Wurden, C. J., {SQUIDs vs. Induction Coils for Ultra-Low Field Nuclear Magnetic Resonance: Experimental and Simulation Comparison}, \emph{IEEE Trans. Appl. Supercond.} 21, 485--468 (2011).







\bibitem{Ledbetter2008PNAS105} Ledbetter, M. P., Savukov, I. M., Budker, D., Shah, V., Knappe, S., Kitching, J., Michalak, D. J., Xu, S. and Pines, A., Zero-field remote detection of NMR with a microfabricated atomic magnetometer, \emph{Proc. Natl. Acad. Sci. USA} 105, 2286-2290, (2008).

\bibitem{Shah2007NatPhot1} Shah, V., Knappe, S., Schwindt, P. D. D. and Kitching, J., Subpicotesla atomic magnetometry with a microfabricated vapour cell, \emph{Nat. Photon.} 1, 649-652, (2007).

\bibitem{Martinez2014NatCommun5} Jimin\'{e}z-Martinez, R., Kennedy, D. J., Rosenbluh, M., Donley, E. A., Knappe, S., Seltzer, S. J., Ring, H. L., Bajaj, V. S. and Kitching, J., Optical hyperpolarization and {NMR} detection of 129Xe on a microfluidic chip, \emph{Nature Commun.} 5, 3908 (2014).






\bibitem{Savukov2009JMR199} Savukov, I. M., Zotev, V. S., Volegov, P. L., Espy, M. A., Matlashov, A. N., Gomez, J.J. and Kraus Jr., R. H., MRI with an atomic magnetometer suitable for practical imaging applications, \emph{J. Magn. Reson.} 199, 188-191 (2009).

\bibitem{Ganssle2014ANIE53} Ganssle, P. J., Shin, H. D., Seltzer, S. J., Bajaj, V. S., Ledbetter, M. P., Budker, D., Knappe, S., Kitching, J. and Pines, A., Ultra-low-field NMR relaxation and diffusion measurements using an optical magnetometer, \emph{Angew. Chem. Int. Edn.} 53, 9766-9770 (2014).

\bibitem{Yashchuk2004PRL93} Yashchuk, V. V., Granwehr, J., Kimball, D. F., Rochester, S. M., Trabesinger, A. H., Urban, J. T., Budker, D. and Pines, A., Hyperpolarized Xenon Nuclear Spins Detected by Optical Atomic Magnetometry, \emph{Phys. Rev. Lett.} 93, 160801, (2004).

\bibitem{Wilzewski2017} Wilzewski, A. Afach, S., Blanchard, J. W. and Budker, D., A Method for Measurement of Spin-Spin Couplings with sub-mHz Precision Using Zero- to Ultralow-Field Nuclear Magnetic Resonance, arXiv:1702.04297 (2017).


\bibitem{Schwindt2010Sandia} Schwindt, P. D. D. and Johnson, C. N., Atomic magnetometer for human magnetoencephalography, Sandia National Laboratories report SAND2010-8443 (2010). \url{http://prod.sandia.gov/techlib/access-control.cgi/2010/108443.pdf}

\bibitem{Baker2006PRL97}  Baker, C. A., Doyle, D. D., Geltenbort, P., Green, K., van der Grinten, M. G. D., Harris, P. G., Iaydjiev, P., Ivanov, S. N., May, D. J. R., Pendlebury, J. M., Richardson, J. D., Shiers, D. and Smith, K. F., Improved Experimental Limit on the Electric Dipole Moment of the Neutron, \emph{Phys. Rev. Lett.} 97, 131801 (2006).






\bibitem{Maul2016RSI87} Maul, A., Bl\"{u}mler, P., Heil, W., Nikiel, A., Otten, E., Petrich, A. and Schmidt, T., Spherical fused silica cells filled with pure helium for nuclear magnetic resonance magnetometry, \emph{Rev. Sci. Instrum.} 87, 015103 (2016).

\bibitem{Kornack2005PRL95} Kornack, T. W., Ghosh, R. K. and Romalis, M. V., Nuclear spin gyroscope based on an atomic magnetometer, \emph{Phys. Rev. Lett.} 95, 230801 (2005).

\bibitem{Donley2010IEEEsensCon} Donley, E. A., Nuclear magnetic resonance gyroscopes, \emph{IEEE Sensors Conf.} 2010, 17-22.




\bibitem{Appelt2010PRL81} Appelt, S., H\"{a}sing, F. W., Sieling, U., Gordji Nejad, A.,Gl\"{o}ggler, S. and Bl\"{u}mich, B., Paths from weak to strong coupling in NMR, \emph{Phys. Rev. A} 81, 023420 (2010).



\bibitem{Bernarding2006JACS128} Bernarding, J., Buntkowsky, G., Macholl, S., Hartwig,  S., Burghoff, M. and Trahms, L., J-coupling nuclear magnetic resonance spectroscopy of liquids in nT fields, \emph{J. Am. Chem. Soc.} 128, 714-715 (2006). 

\bibitem{Bevilacqua2016JMR263} Bevilacqua, G., Biancalana, V., Baranga, A. B.-A., Dancheva, Y. and Rossi, C., Microtesla NMR J-coupling spectroscopy with an unshielded atomic magnetometer, \emph{J. Magn. Reson.} 263, 65-70 (2016). 

\bibitem{Trahms2010MRI28} Trahms, L. and Burghoff, M., {NMR} at very low fields, \emph{Magn. Reson. Imag.} 28, 1244-1250 (2010). 

\bibitem{Shim2014JMR246} Shim, J. H., Lee, S.-J., Hwang, S.-m., Yu, K.-K. and Kim, K., Two-dimensional {NMR} spectroscopy of methanol at less than 5 microtesla, \emph{J. Magn. Reson.} 246, 4-8 (2014).





\bibitem{Thiel2007RSI78} Thiel, F., Schnabel, A., Knappe-Gr\"{u}neberg, S., Stollfu\ss, D. and Burghoff, M., Demagnetization of magnetically shielded rooms, \emph{Rev. Sci. Instrum.} 78, 035106 (2007).


\bibitem{Biancalana2014RSI85} Biancalana, V., Dancheva, Y. and Stiaccini, Note: a fast pneumatic sample-shuttle with attenuated shocks, \emph{Rev. Sci. Instrum.}, 85, 036104 (2014).










\bibitem{Bowers1986PRL57} Bowers, C. R. and Weitekamp, D. P., Transformation of symmetrization order to nuclear-spin magnetization by chemical reaction and nuclear magnetic resonance, \emph{Phys. Rev. Lett.}, 57, 2645-2648 (1986). 

\bibitem{Bowers1987JACS109} Bowers, C. R. and Weitekamp, D. P., Parahydrogen and synthesis allow dramatically enhanced nuclear alignment, \emph{J. Am. Chem. Soc.}, 137, 5541-5542 (1987). 

\bibitem{Adams2009Science323} Adams, R. W., Aguilar, J. A., Atkinson, K. D., Cowley, M. J., Elliott, P. I., Duckett, S. B., Green, G. G., Khazal, I. G., Lopez-Serrano, J. and Williamson, D. C., Reversible interactions with para-hydrogen enhance {NMR} sensitivity by polarization transfer, \emph{Science}, 323, 1708-1711 (2009). 

\bibitem{Colell2017JPCC} Colell, J. F. P., Logan, A. J. W., Zhou, Z., Shchepin, R. V., Barskiy, D. A., Ortiz Jr., G. X., Wang, Q., Malcolmson, S. J., Chekmenev, E. Y., Warren, W. S., and Theis, T., Generalizing, Extending, and Maximizing Nitrogen-15 Hyperpolarization Induced by Parahydrogen in Reversible Exchange
\emph{J. Phys. Chem. C}, Article ASAP
DOI: 10.1021/acs.jpcc.6b12097 (2017)

\bibitem{Theis2015JACS137} Theis, T., Truong, M. L., Coffrey, A. M., Shchepin. R. V., Waddell, K. W., Shi, F., Goodson, B. M., Warren, W. S. and Chekmenev, E. Y., Microtestla {SABRE} enables $10\%$ nitrogen-15 nuclear spin polarization, \emph{J. Am. Chem. Soc.}, 137, 1404-1407 (2015). 

\bibitem{Zhivonitko2015CCommum51} Zhivonitko, V. V., Skovpin, I. V. and Koptyug, I. V, Strong $^{31}$P nuclear spin hyperpolarization produced via reversible chemical interaction with parahydrogen, \emph{Chem. Commun.}, 51, 2506-2509 (2015).

\bibitem{Theis2011Nature473} ``Spectroscopy: {NMR} without the magnet'', Research highlight in \emph{Nature} 473, 126 (2011) doi:10.1038/473126b

\bibitem{Theis2011NatPhys7} Theis, T., Ganssle, P., Kervern, G., Knappe, S., Kitching, J., Ledbetter, M. P., Budker, D. and Pines, A., Parahydrogen-enhanced zero-field nuclear magnetic resonance, \emph{Nat. Phys.}, 7, 571-575 (2011). 


\bibitem{Butler2013JCP138} Butler, M., Kervern, G., Theis, T., Ledbetter, M. P., Ganssle, P. J., Blanchard, J. W., Budker, D. and Pines, A., Parahydrogen-induced polarization at zero magnetic field, \emph{J. Chem. Phys.} 138, 234201 (2013).

\bibitem{Theis2012JACS134} Theis, T., Ledbetter, M. P., Kervern, G., Blanchard, J. W., Ganssle, P. J., Butler, M., Shin, H. D., Budker, D. and Pines, A., Zero-field NMR enhanced by parahydrogen in reversible exchange, \emph{J. Am. Chem. Soc.}, 134, 3987-3990 (2012). 




\bibitem{Lee1987JMR75} Lee, C. J., Suter, D. and Pines, A., Theory of multiple-pulse {NMR} at low and zero fields, \emph{J. Magn. Reson.} 75, 110-124 (1987).

\bibitem{Llor1991PRL67} Llor, A., Olejniczak, Z., Sachleben, J. and Pines, A., Scaling and time reversal of spin couplings in zero-field {NMR}, \emph{Phys. Rev. Lett.} 67, 1989-1992 (1991).

\bibitem{Sjolander2017JPCLXX} Sjolander, T. F., Tayler, M. C. D., Kentner, A., Budker, D. and Pines, A., Homonuclear $J$-coupling spectroscopy using two-dimensional nuclear magnetic resonance at zero field, \emph{J. Phys. Chem. Lett.} in press (2017).




\bibitem{Tayler2016JMRXX} Tayler, M. C. D., Sjolander, T. F., Pines, A. and Budker, D., Nuclear magnetic resonance at millitesla fields using a zero-field spectrometer, \emph{J. Magn. Reson.} 270, 35-39 (2016).

\bibitem{Sjolander2016JPCAXX} Sjolander, T. F., Tayler, M. C. D., Budker, D. and Pines, A., Transition-selective pulses in zero-field nuclear magnetic resonance, \emph{J. Phys. Chem. A} 120, 4343-4348 (2016).






\bibitem{Hilborn1982AJP50} Hilborn, R. C., Einstein coefficients, cross-sections, f values, dipole moments and all that, \emph{Am. J. Phys.} 50, 982 (1982).

\bibitem{Hilborn1982AJP50err} Hilborn, R. C., Erratum: ``Einstein coefficients, cross-sections, f values, dipole moments and all that'', \emph{Am. J. Phys.} 51, 471 (1983).

\bibitem{Macaluso18981} Macaluso, D. and Corbino, O. M., Sopra Una Nuova Azione Che La Luce Subisce Attraversando Alcuni Vapori Metallici In Un Campo Magnetico, \emph{Nuovo Cimento} 8, 257 (1898).

\bibitem{Macaluso18982} Macaluso, D. and Corbino, O. M., Sulla Relazione Tra Il Fenomeno Di Zeemann E La Rotazione Magnetica Anomala Del Piano Di Polarizzazione Della Luce, \emph{Nuovo Cimento} 9, 384 (1898).

\bibitem{Zeeman1897Nature55} Zeeman, P., The effect of magnetisation on the nature of light emitted by a substance, \emph{Nature} 55, 347 (1897).

\bibitem{Happer1972RMP44} Happer, W., Optical pumping, \emph{Rev. Mod. Phys.}, 44, 169-249, (1972).

\bibitem{Happer1967PR163} Happer, W. and Mathur, B. S.; Effective Operator Formalism in Optical Pumping \emph{Phys. Rev.} 163, 12-25 (1967).

\bibitem{MolischOehrybook} Molisch, A. F. and Oehry, B. P., Radiation trapping in atomic vapors, Oxford Science Publications, Clarendon Press, Oxford (1998). ISBN 0198538669.

\bibitem{SteckRb201} Steck, D. A., Rubidium-87 D-line data, online at, \url{http://steck.us/alkalidata} revision 2.1.5 (2015).




\bibitem{Xu2008JMRI28} Xu, S., Harel, E., Michalak, D. J., Crawford, C. W., Budker, D. and Pines, A., Flow in porous metallic materials: a magnetic resonance imaging study, \emph{J. Magn. Reson. Imag.} 28, 1299-1302 (2008). 

\bibitem{King2016XX} King, J. P., Sjolander, T. F., Blanchard, J. W., Antisymmetric couplings enable direct observation of chirality in nuclear magnetic resonance spectroscopy, \emph{J. Phys. Chem. Lett.}, accepted manuscript \url{http://pubs.acs.org/doi/abs/10.1021/acs.jpclett.6b02653} (2016).

\bibitem{Abragam1978RPP41} Abragam, A. and Goldman, M., Principles of dynamic nuclear polarisation, \emph{Rep. Prog. Phys.} 41, 395 (1978). 

\bibitem{WenckebachBOOK2016} Wenckebach, T., \emph{Essentials of dynamic nuclear polarization}, Spindrift Publications; 1st edition (2016).  ISBN: 978-9075541182.

\bibitem{Kockenberger2014emagres} K\"{o}ckenberger, W., Dissolution dynamic nuclear polarization, \emph{eMagRes} 3, 161-170 (2014). DOI:10.1002/9780470034590.emrstm1311

\bibitem{Lovchinsky2016Science351} Lovchinsky, I., Sushkov, A. O., Urbach, E., de Leon, N. P., Choi, S., De Greve, K., Evans, R., Gertner, R., Bersin, E., M{\"u}ller, C., McGuinness, L., Jelezko, F., Walsworth, R. L., Park, H. and Lukin, M. D., Nuclear magnetic resonance detection and spectroscopy of single proteins using quantum logic, \emph{Science}, 351, 836-841, (2016).

\bibitem{Xu2006PNAS103} Xu, S., Yashchuk, V. V., Donaldson, M. H., Rochester, S. M., Budker, D. and Pines, A., Magnetic resonance imaging with an optical atomic magnetometer, \emph{Proc. Natl. Acad. Sci. USA} 103, 12668-12671 (2008). 

\bibitem{Xu2008PRA78} Xu, S., Crawford, C. W., Rochester, S. M., Yashchuk, V. V., Budker, D. and Pines, A., Submillimeter-resolution magnetic resonance imaging at the Earth's magnetic field with an atomic magnetometer, \emph{Phys. Rev. A} 78, 013404 (2006).

\bibitem{Savukov2013JMR231} Savukov, I. and Karaulanov, T., Anatomical MRI with an atomic magnetometer, \emph{J. Magn. Reson. Imag.} 231, 39-45 (2013). 

\bibitem{Savukov2013JMR233} Savukov, I., Karaulanov, T., Wurden, C. J. V. and Schultz, L. Non-cryogenic ultra-low field MRI of wrist-forearm area, \emph{J. Magn. Reson. Imag.} 233, 103-106 (2013). 

\bibitem{Clarke2007ARBE9} Clarke, J., Hatridge, M. and M\"{o}{\ss}le, M., {SQUID-detected magnetic resonance imaging in microtesla fields}, \emph{Annu. Rev. Biomed. Eng.} 9, 389-413 (2007). 


\end{thebibliography}
\end{document}